\documentclass[preprint,12pt]{elsarticle}

\usepackage{amssymb,amsmath,amsthm,mathtools,amsbsy,amsfonts}
\usepackage[margin=1.in]{geometry}
\usepackage{caption}
\usepackage{subcaption}
\usepackage{float}
\usepackage{graphicx}
\usepackage{multicol}
\usepackage{lipsum}
\usepackage[dvipsnames]{xcolor}
\newcommand{\R}{\mathbb{R}}
\newcommand{\p}{\partial}
\newcommand{\Div}{\text{div}}
\newcommand{\pLap}{\Delta_p}
\newcommand{\pLapS}{\Delta_p^S}
\newcommand{\grad}{\nabla}
\newcommand{\gradS}{\nabla^S}
\newcommand{\heatmethod}{heat method }
\newcommand{\poissonmethod}{$p$-Poisson method }
\usepackage{hyperref}
\usepackage{cleveref}

\DeclareMathOperator*{\argmin}{arg\,min}

\journal{Computer Aided Geometric Design}

\begin{document}

\begin{frontmatter}



\title{Geodesic distance approximation using a surface finite element method for the $p$-Laplacian}


\author[inst1]{Hannah Potgieter}
\author[inst1]{Razvan C. Fetecau}
\author[inst1]{Steven J. Ruuth}

\affiliation[inst1]{organization={Department of Mathematics, Simon Fraser University},
            city={Burnaby},
            province={British Columbia},
            country={Canada}}

\begin{abstract}
We use the $p$-Laplacian with large $p$-values in order to approximate geodesic distances to features on surfaces. This differs from Fayolle and Belyaev's (2018) \cite{FAYOLLE20181} computational results using the $p$-Laplacian for the {\em distance-to-surface} problem. Our approach appears to offer some distinct advantages over other popular PDE-based distance function approximation methods. We employ a surface finite element scheme and demonstrate numerical convergence to the true geodesic distance functions. We check that our numerical results adhere to the triangle inequality and examine robustness against geometric noise such as vertex perturbations. We also present comparisons of our method with the heat method from Crane et al. \cite{Crane} and the classical polyhedral method from Mitchell et al. \cite{Mitchell}.

\end{abstract}


\end{frontmatter}



\section{Introduction} 
\label{sec:intro}
Problems involving the $p$-Laplace operator appear in a variety of applications including data clustering (Bühler and Hein \cite{BuhlerHein}), image processing (Blomgren et al. \cite{Blomgren}; Chen et al. \cite{Chen2006}, Caselles et al. \cite{Caselles}), distance function approximation (Fayolle \& Belyaev \cite{FAYOLLE20181}), optimal transport (Fayolle \& Belyaev \cite{FAYOLLE20181}), non-Newtonian fluids (Ružička \cite{Ruzicka}), and shape morphing (Cong et al. \cite{shapeMet}). In this paper, we focus on the application of the $p$-Laplacian for computing geodesic distances on surfaces. 

Bhattacharya et al. \cite{bhattacharya1989limits} consider the following boundary value problem (BVP) for the $p$-Laplacian in a bounded domain $\Omega \subset \R^n$ with boundary $\p \Omega$:
\begin{subequations} \label{eq:1}
\begin{alignat}{2}
\label{eq:1a}  -\pLap u_p &= 1, \quad \, \text{ in } \Omega,  \\[3pt]
\label{eq:1b} u_p &= 0, \qquad  \text{ on }  \p \Omega
\end{alignat}
\end{subequations}
for $2 \leq p \leq \infty$. The $p$-Laplacian is defined by $\pLap u = \Div \left( |\grad u |^{p-2} \grad u \right)$; note that for $p=2$ it reduces to the usual Laplacian. Bhattacharya et al. \cite{bhattacharya1989limits} show that the solution $ u_p(x)$ to the BVP \eqref{eq:1a}-\eqref{eq:1b} converges strongly as $p \to \infty$, in a certain functional sense, to $\mathrm{dist}(x, \p \Omega)$, where $\mathrm{dist}(\cdot,\p \Omega)$ denotes the distance function to the boundary $\p \Omega$. The problem \eqref{eq:1a}-\eqref{eq:1b} is fully nonlinear and degenerate, unless $p = 2$ in which case it is a linear Poisson problem. Taking $p = 2$ yields what is referred to as the `Poisson distance' and $p > 2$ gives the `$p$-Poisson distance'. 

Fayolle \& Belyaev \cite{FAYOLLE20181} use \eqref{eq:1a}-\eqref{eq:1b} to compute the {\em distance-to-surface} problem, where $\Omega$ is a subset of $\R^3$ such that $\p \Omega$ represents the surface of interest. In this paper, we are interested in the more general {\em distance-to-feature} problem on {\em surfaces} both with and without boundaries. A `feature' can be any collection of points or curves on a given surface, to which we want to compute the distance. Such points or curves may lie either in the interior or on the boundary of the surface. We point out that our interest is to compute the surface geodesic (intrinsic) distance function to such features. This is in contrast with Fayolle \& Belyaev \cite{FAYOLLE20181} who compute the extrinsic distance (in $\R^3$) to a given surface. 

Computer graphics problems frequently call for computation of distances to features on a discrete surface. The distance-to-feature problem is practically useful for a variety of tasks such as segmentation, deformation, and path planning as mentioned in \cite{FAYOLLE20181}, \cite{Lipman}, and \cite{Solomon}. It is also useful for medical image analysis (see Naber et al. \cite{NaberMedImage}). Belyaev \& Fayolle \cite{Belyaev2015} applied alternating direction method of multipliers (ADMM) schemes to different energy minimization problems seeking to achieve accurate distance function approximations, and showed in \cite{FAYOLLE20181} that minimizing the $p$-Poisson energy yields the best results. Accordingly, we extend their work with the $p$-Laplacian to the distance-to-feature problem and perform more comprehensive numerical tests.  

\subsection{Related works}
\label{sec:relwork}
The problem of approximating geodesic distances on surfaces has been extensively studied, with a variety of approaches proposed. See, for example, \cite{Crane}, \cite{Lipman}, \cite{Solomon},\cite{Belyaev2015}, \cite{WANG2017262}, and \cite{KimmelSethian1998}. These range from PDE-based methods to computational geometry techniques, each offering distinct advantages and limitations. Among these, the {\it \heatmethod}of Crane, Weischedel, and Wardetzky \cite{Crane} and the {\it polyhedral method} of Mitchell, Mount, and Papadimitriou \cite{Mitchell} are two prominent approaches that we utilize  for comparison.

The \heatmethod \cite{Crane}, which computes geodesic distances using Varadhan’s formula \cite{Varadhan}, is one of the most well-known PDE-based approaches. The \heatmethod splits the problem into first finding the direction along which distance is increasing and then computing the distance itself. The method is computationally efficient, adaptable, robust against noisy surface data, and simple.  However, the \heatmethod has notable limitations. It struggles with boundary conditions on open surfaces, such as hemispheres, an issue observed in our experiments and also seen in Crane et al.'s experiments \cite{Crane}. Additionally, the method often fails to satisfy the triangle inequality and produces smooth approximations \cite{Solomon}. Concurrent with our work, Feng and Crane \cite{FengCrane} introduced the Signed Heat Method (SHM) for computing generalized signed distance functions. By formulating vector heat diffusion with homogeneous Neumann conditions directly on the vector field, rather than on scalar potentials, SHM yields suitable distance approximations at surface boundaries without requiring heuristics or post-processing corrections. Our comparisons are performed against the (scalar) heat method \cite{Crane}.

The polyhedral method \cite{Mitchell}, in contrast, operates on polyhedral surfaces and employs a ``continuous Dijkstra technique,” which adapts Dijkstra’s algorithm \cite{Dijkstra1959} from graph structures to polyhedral surfaces. By carefully incorporating geometric considerations, the method computes exact geodesics along polyhedral surfaces. It is typically both computationally efficient and highly accurate when applied to tessellated surfaces. We include this method in our comparisons to better understand the properties of our solutions, and
refer to a recent survey paper \cite{crane2020survey} for further details on the method, its attributes and limitations.

Other notable approaches include Sethian’s fast marching method \cite{Sethian1996} for numerical approximation of solutions to the Eikonal equation $|\nabla u(x)| = 1$, Solomon et al.'s \cite{Solomon} optimal transport inspired framework, Edelstein et al.'s \cite{ConvexNew} PDE-based convex optimization approach, as well as graph-based methods, such as those proposed by Adikusuma et al. \cite{Adikusuma}.

In this paper, we present a new algorithm for geodesic distance computation and evaluate its key properties. As part of this analysis, we compare our approach with the \heatmethod and the polyhedral method, focusing on illustrative examples to highlight the key properties of our approach. We do not aim to provide an exhaustive comparison in these tests, nor do we compare with other well-established methods. For a broader overview of geodesic distance approximation techniques and their respective merits, see \cite{crane2020survey}. 

\subsection{Contributions}
\label{sec:contrib}
This paper introduces a new method for approximating intrinsic geodesic distances to features on surfaces. This {\it\poissonmethod}is formulated using a 
$p$-Laplace equation intrinsic to the surface, employing appropriate mixed boundary conditions to define the feature sets. In the large 
$p$ limit, the solution to the equations provides the solution to the intrinsic geodesic problem.

In our implementation, the Alternating Direction Method of Multipliers (ADMM) is employed alongside a surface finite element method with piecewise linear elements to solve the mixed boundary value problem. Convergence of the method is assessed through numerical studies on the hemisphere and torus, as well as on more complex surface meshes, with comparisons to the \heatmethod and the polyhedral method. We find convergence to the desired intrinsic geodesic distance. Additionally, our proposed method provides an improved treatment of boundaries relative to the \heatmethod in certain examples. Robustness tests further demonstrate the preservation of the triangle inequality and stability against geometric perturbations.

The aim of this work is not to compete with state-of-the-art geodesic distance approximation methods on surface meshes, which are generally faster, especially in high-accuracy scenarios. However, our PDE formulation does not require a mesh and offers a path to implementation for other surface representations, such as level sets or point clouds, through suitable numerical PDE techniques, including mesh-free methods in the case of point clouds. As well, this work provides insights into the behavior of the surface
$p$-Laplacian for large $p$.

The method has certain limitations. Most notably, it is slower than the polyhedral method for surfaces defined on meshes, particularly when highly accurate solutions are required. Additionally, suitable solvers must be chosen to accommodate the surface representation and its regularity. Our implementation focuses on triangle and quadrilateral surface meshes using the well-known FEM package deal.II \cite{dealII94} coupled with ADMM; other surface representations and solvers are not explored in this work. Finally, we note that correctness is primarily assessed through numerical experiments. 

\section{Problem formulation} 
\label{sec:model}

\begin{figure}
    \centering
    \includegraphics[scale = 0.3]{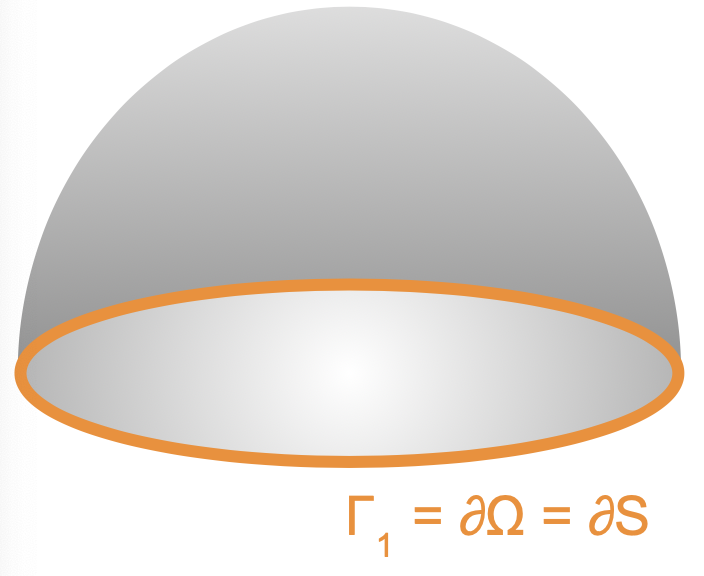}\\
    \includegraphics[scale = 0.3]{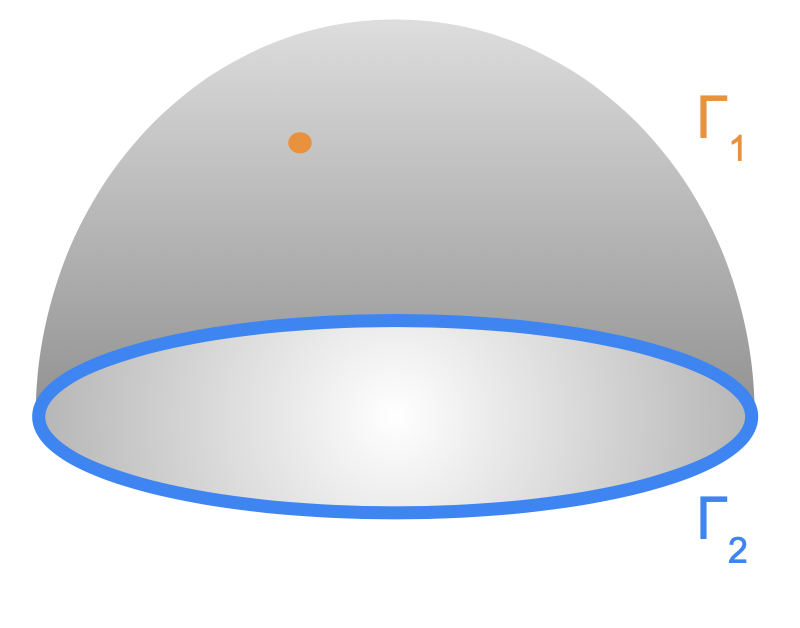}
    \includegraphics[scale = 0.3]{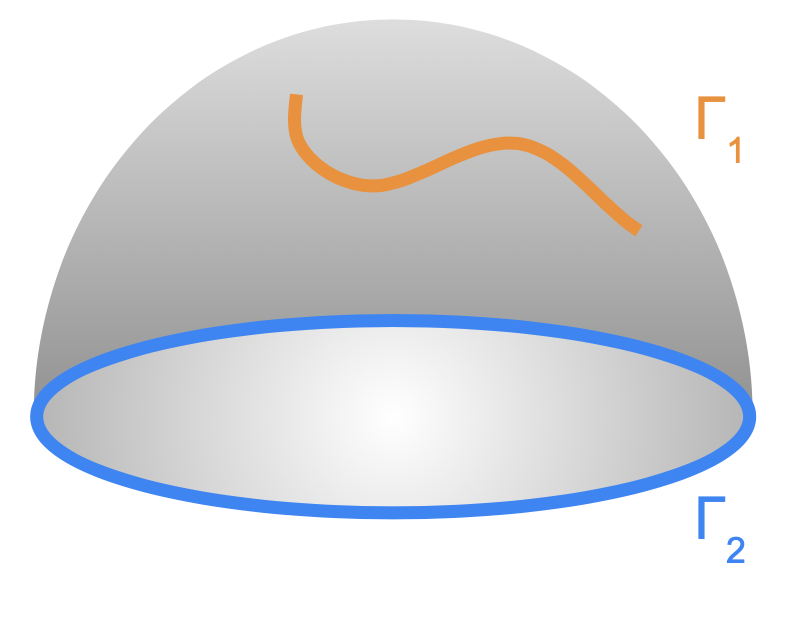}
    \includegraphics[scale = 0.3]{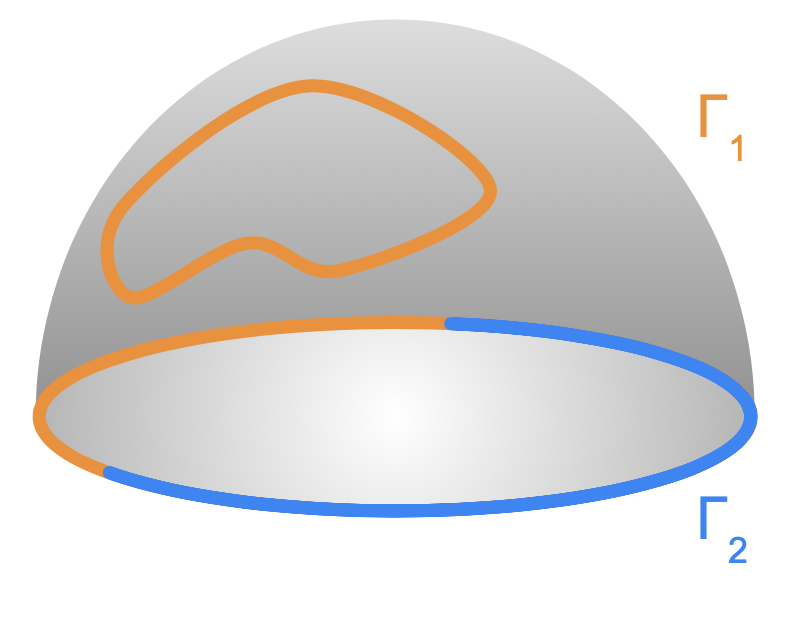}
    \caption{$S$ is the open hemisphere and $\Gamma_1$ (indicated in orange) represents the feature set. Top row:  Distance-to-boundary  problem ($\Gamma_1= \p S, \Gamma_2 = \emptyset$). Bottom row: Distance-to-feature problem. $\Gamma_2$ is indicated in blue.}
    \label{fig:setup}
\end{figure}

Consider a generic surface $S$ and the $p$-Laplacian operator on $S$ given by
\begin{equation*}
     \pLapS u = \Div^S \Bigl( |\gradS u |^{p-2} \gradS u \Bigr),
\end{equation*}
where $\Div^S$ and $\gradS$ denote the surface divergence and gradient, respectively. Let $\Gamma_1 \subset S \cup \p S$ denote the `feature' set, i.e., the set to which we want to compute the geodesic distance. The set $\Gamma_1$ may consist of any collection of points or curves from the interior or boundary of $S$. Denote the open set $\Omega$ by $\Omega = S \setminus \Gamma_1 $ and $\Gamma_2 = \p S \setminus \Gamma_1$; note that $\p \Omega = \Gamma_1 \cup \Gamma_2$, where the union is disjoint. \Cref{fig:setup} illustrates examples of $\Gamma_1$ and $\Gamma_2$ on the open hemisphere $S$. Note that $\Gamma_1$ may consist of points, closed curves, and open curves, and may also overlap with the natural surface boundary $\p S$. 

For $2 \leq p \leq \infty$ fixed, we consider the following BVP for the surface $p$-Laplacian:
\begin{subequations}
\begin{alignat}{3}
\label{eq:2a}  -\pLapS u_p &= 1, \quad \, \text{ in } \Omega,  \\[3pt]
\label{eq:2b} u_p &= 0, \qquad \text{ on } \Gamma_1 \subseteq \p \Omega,  \\[3pt]
\label{eq:2c}   \frac{\p u_p}{\p n} & = 0, \qquad \text{ on } \Gamma_2 \subseteq \p \Omega.
\end{alignat}
\end{subequations}
Note that Dirichlet conditions are imposed on $\Gamma_1$ (the feature set) and Neumann conditions are imposed on $\Gamma_2$, which represents the unconstrained portion of the surface boundary. 

The $p$-Poisson problem can be derived from the minimization of the variational integral
\begin{equation}
    E_p(u) \equiv \frac1p \int_\Omega  |\gradS u|^p dA - \int_\Omega u \, dA.
    \label{eq:Ep}
\end{equation}
Setting $0 < p < 1$ would lead to a non-convex energy, however, we are interested in the range $p \in [2, \infty)$. The Euler-Lagrange equations for the functional \eqref{eq:Ep} give rise to equation \eqref{eq:2a}. We enforce the boundary condition 
\eqref{eq:2b}, as the distance from the feature set to itself must be zero. This leaves \eqref{eq:2c} to be justified. To this purpose we use the notion of {\em natural} boundary conditions in calculus of variations \cite{Stein} as is done by Edelstein et al. \cite{ConvexNew}. Note that there are no specific boundary conditions on $\Gamma_2$ which need to be imposed beforehand, by the distance-to-feature problem that we consider. Hence, on the unconstrained portion $\Gamma_2$ of $\partial \Omega$, we impose the natural boundary condition, which achieves the lowest energy among all possible boundary conditions \cite{Stein}. 

By a standard calculation, the natural boundary condition corresponding to the energy (\ref{eq:Ep}) (with \eqref{eq:2b} already imposed), is given by
\begin{equation}
    |\gradS u|^{p-2} (\gradS u \cdot \mathbf{n}) = 0 \quad \text{on } \Gamma_2,
    \label{eq:naturalBC}
\end{equation}
where $\mathbf{n}$ is the outward unit normal to $\partial \Omega$.  In order to satisfy (\ref{eq:naturalBC}), we need to have either $|\gradS u|^{p-2} = 0$ or $\gradS u \cdot \mathbf{n} = 0$ for all points on $\Gamma_2 \subset \partial \Omega$. We are interested in a distance function approximation, so as $p \to \infty$ it is desired that $|\gradS u_p|$ will be close to $1$. Hence, the homogeneous Neumann boundary condition \eqref{eq:2c} follows. Below, we present further mathematical justification for the mixed BVP \eqref{eq:2a}-\eqref{eq:2c}, and at the end of this section, we go into more detail for a particular example in $1$D.

The particular case when $\Gamma_1 = \p \Omega$, $\Gamma_2 = \emptyset$, and $\Omega$ is a bounded domain in $\mathbb{R}^n$, was considered in the mathematical literature by Bhattacharya et al. \cite{bhattacharya1989limits},  and in the computational work on distance approximation of Fayolle and Belyaev \cite{FAYOLLE20181}. In this restrictive setting, problem \eqref{eq:1a}-\eqref{eq:1b} computes (in the limit $p \to \infty$) the distance in the Euclidean space to the boundary of a set $\Omega \subset \R^n$. In an extension of the work from \cite{bhattacharya1989limits}, \cite{Azorero_etal2009-mlimits} considered the more general case when $\Gamma_1$ is a strict subset of $\p \Omega$ (hence, $\Gamma_2 \neq \emptyset$). Specifically, the authors of \cite{Azorero_etal2009-mlimits} considered the mixed BVP \eqref{eq:2a}-\eqref{eq:2c} on smooth convex subsets of $\mathbb{R}^n$, and investigated the limit $p\to \infty$ of its weak solutions $u_p$. They proved (see \cite[Remark 2.2]{Azorero_etal2009-mlimits}) that 
\begin{equation}
\label{eqn:limit-p}
\lim_{p \to \infty} u_p(x) = \mathrm{dist} (x, {\Gamma_1}).
\end{equation}
The convergence is, in fact, uniform in $\overline{\Omega}$. From an analysis point of view, the approach taken in \cite{Azorero_etal2009-mlimits} to study the $p\to \infty$ limit, is very similar to that from \cite{bhattacharya1989limits}. The reason the methods extend 
with very few modifications from \cite{bhattacharya1989limits} (where $\Gamma_2 = \emptyset$) to the mixed BVP with $\Gamma_2 \neq \emptyset$, lies exactly in the fact that the homogeneous Neumann BC \eqref{eq:2c} is natural \cite{Stein}. 

In the present work, we are interested in geodesic distance approximation on surfaces. For this reason, the BVP \eqref{eq:2a}-\eqref{eq:2c} is set up on a subset $\Omega$ of a surface $S$ using the surface $p$-Laplacian, and we compute the geodesic distance function to the feature set $\Gamma_1 \subset \p \Omega$. The feature set can be very diverse, as illustrated in \Cref{fig:setup}; $\Gamma_1$ may or may not have any points on the surface's boundary $\p S$. When $\Gamma_1 = \p S$ (see top row in \Cref{fig:setup}), then $\Omega = S$, and the distance-to-feature problem simplifies to the distance-to-boundary problem as in Bhattacharya et al. \cite{bhattacharya1989limits}, except that here we are interested in the geodesic, rather than Euclidean, distance.
For numerical purposes, large $p$-values are of interest. Unfortunately, large $p$-values lead to ill-conditioned problems and are therefore numerically challenging to solve \cite{Huang2007PreconditionedDA}. 

We further note that the analysis in \cite{bhattacharya1989limits, Azorero_etal2009-mlimits} can be extended to compact Riemannian manifolds using the Sobolev and Kondrachov embedding theorems on manifolds from \cite{Aubin-book}. Hence, one can prove rigorously the limit \eqref{eqn:limit-p} in general surface settings, as considered in our work. Also, from a PDE analysis point of view, the case when $\Gamma_1$ contains isolated points (see left plot in bottom row of  \Cref{fig:setup}), is degenerate. In this case, one can consider $\Gamma_1$ to be the boundary of an arbitrarily small geodesic ball around the single point.  

\paragraph{Exact 1D solution} 

For concreteness, we show a simple 1D example with an exact solution. A similar calculation may be performed for a radially symmetric problem in 2D, however solutions are not as simple and involve hypergeometric functions. Consider $S=(-1,1)$ with feature set $\Gamma_1 = \{0\}$. Then, $\Omega = (-1, 0) \cup (0, 1)$ and $\Gamma_2 = \{ -1, 1\} $. The exact solution for this problem is
\[
\mathrm{dist} (x,  \Gamma_1) = |x|. 
\]

The one-dimensional $p$-Laplacian is given by $\pLap u_p(x) =  \left (|u'|^{p-2} u' \right )'$, and the BVP problem \eqref{eq:2a}-\eqref{eq:2c} becomes 
\begin{subequations}
\begin{alignat}{3}
\label{eq:a'} -\left (|u'|^{p-2} u' \right )' &= 1,  \quad  0 < |x| <1,  \\[3pt]
\label{eq:b'} u_p(0) &= 0,  \\[3pt]
\label{eq:c'}   u_p'(-1) &=  u_p'(1) = 0. 
\end{alignat}
\end{subequations}

The exact solution of \eqref{eq:a'}-\eqref{eq:c'} is
\begin{equation}
\label{eq:1d-exact}
    u_p(x) = -\frac{p-1}{p} \left (1-|x| \right )^{\frac{p}{p-1}} +  \frac{p-1}{p}.
\end{equation}
Using \eqref{eq:1d-exact} one can then show immediately that for any $x \in \Omega$ fixed,
\begin{equation}
\label{eq:limit-exact}
   \lim_{p \to \infty} u_p(x) = |x| =  \mathrm{dist}(x, \Gamma_1). 
\end{equation} 
We note that the convergence \eqref{eq:limit-exact} is in fact uniform on $\Omega$. Indeed, use  
\begin{equation*}
    \frac{\p}{\p x}\bigg[ |x|-u_p(x) \bigg] = \mathrm{sign}(x) \bigg( 1-\left( 1-|x| \right)^{\frac{1}{p-1}} \bigg),
\end{equation*}
for $x \in \Omega$. Since $( 1-\left( 1-|x| \right)^{\frac{1}{p-1}} ) \geq 0$ for all $x \in \Omega$, we infer that $|x|-u_p(x)$ is a monotone function and hence,
\begin{equation*}
\begin{split}
    \Bigl| |x|-u_p(x) \Bigr| & \leq \bigl| |\pm 1|-u_p( \pm 1) \bigr| \leq 1-\frac{p-1}{p} = \frac{1}{p}, \qquad \text{ for all } x \in \Omega.
\end{split}
\end{equation*}
The uniform convergence in \eqref{eq:limit-exact} can then be concluded.

\begin{figure}
    \centering
    \includegraphics[scale = 0.45]{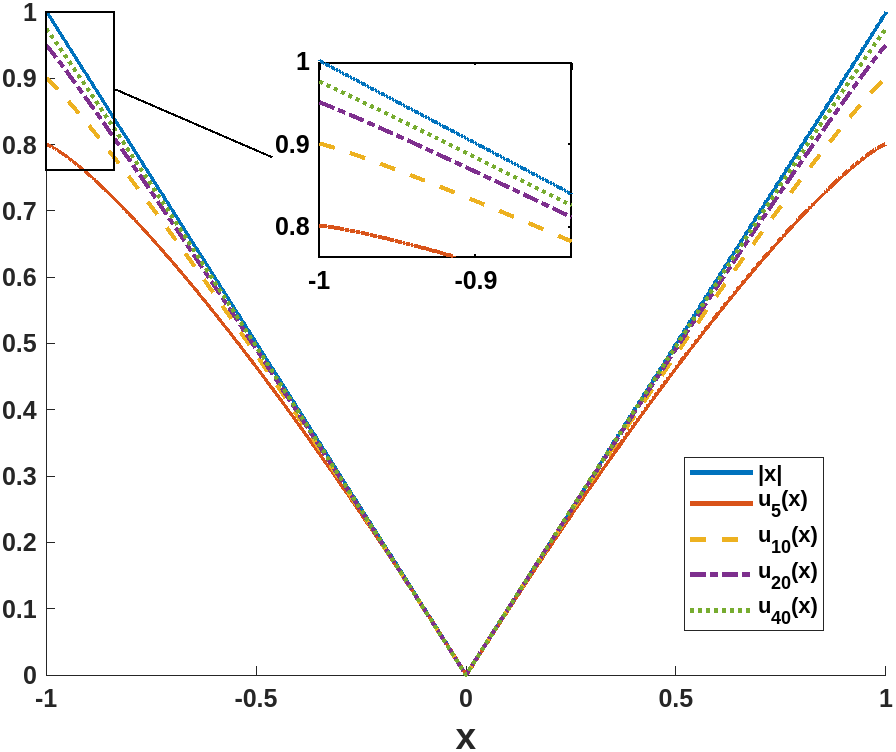}
    \includegraphics[scale = 0.45]{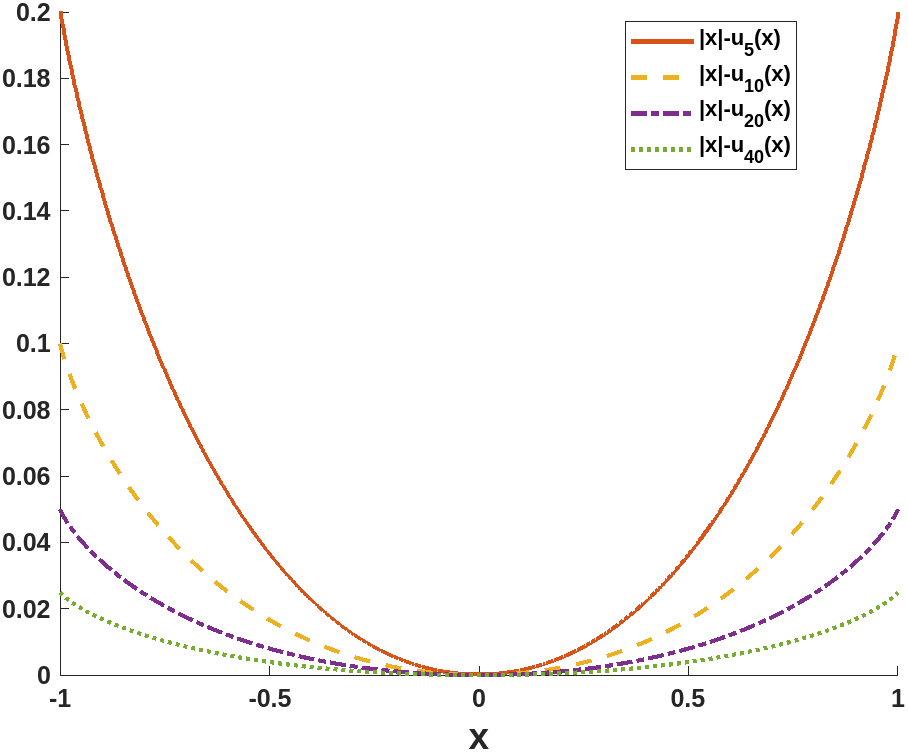}
    \caption{One-dimensional example, with $S=(-1,1)$ and $\Gamma_1=\{0\}$. Left: the exact solution $u_p(x)$ (see \eqref{eq:a'}-\eqref{eq:c'} and \eqref{eq:1d-exact}) converges to $|x|=\mathrm{dist} (x,\Gamma_1)$ as $p$ increases. Right: the corresponding pointwise errors $|x|-u_p(x)$, for various $p$-values.}
    \label{fig:1d}
\end{figure}

\Cref{fig:1d} shows the exact solution \eqref{eq:1d-exact} for various values of $p$, and their corresponding errors, $|x|-u_p(x)$. Here we use \eqref{eq:2c}, ensuring a well-posed problem for our
computations with finite $p$. In the $p \to \infty$ limit, $u_p(x)$ violates the Neumann condition \eqref{eq:c'} (see the zoom-in of the left plot in \Cref{fig:1d}), which is the desired behaviour for the distance problem. Accordingly, in computations with large $p$ the error from the homogeneous Neumann condition \eqref{eq:c'} does not interfere with the linear convergence rate of $1/p$. This supports our choice of homogeneous Neumann boundary conditions on $\Gamma_2$.

\section{Surface finite element method}

\label{sec:surfaceFEM}

The continuous surface $\Omega$ is replaced by a piecewise polynomial surface $\Omega_h$; this introduces a geometric error between $\Omega$ and $\Omega_h$. We will use a finite element space: 
\begin{equation*}
   S_h = \{ \phi_h \in C(\Omega_h) : \phi_h|_T  \text{ is in } P^1 \text{ for each } T \in \mathcal{T}_h\},
\end{equation*}
where $P^1$ is the space of linear elements on $\mathcal{T}_h$, the collection of faces on the surface mesh. 
The corresponding lifted finite element space is
\begin{equation*}
    S_h^l = \{ \varphi_h = \phi_h^l : \phi_h \in S_h\},
\end{equation*}
where the superscript denotes the constant extension from $\Omega$ in the normal direction or lift onto $\Omega$. 

Convergence has been established for the surface finite element method (SFEM) with linear elements. In particular, for the surface Poisson problem, $\Delta^S u = f$, Dziuk \& Elliot \cite{DziukElliott} show that if $ || f-f_h||_{L_2(\Omega)} \leq c_f h^2$, then 
\begin{equation*}
    || u-u_h||_{L_2(\Omega)} \leq c_1 h^2 , \quad || \gradS (u-u_h)||_{L_2(\Omega)} \leq c_2 h, 
\end{equation*}
which is analogous to the Euclidean result.

Distance functions are typically nonsmooth which means we do not see improved convergence rates with higher order elements. Therefore, we consider only first order elements. We choose to use SFEM due to the nonsmooth nature of distance functions. Using a finite difference discretization paired with the closest point embedding method \cite{CPM} suffices for smooth problems. However, finite difference discretizations fail to converge for the examples we consider. The divergence structure of the $p$-Laplacian allows us to employ a variational finite element method which is more robust against nonsmooth data. 

\paragraph{Alternating direction method of multipliers (ADMM)}

To solve the nonlinear $p$-Poisson problem, we use two methods: a standard Newton scheme and ADMM. Note that throughout the paper, we will present only the results obtained with ADMM. Using the Newton scheme we were able to obtain very similar results; however, Newton is less robust for large $p$ and requires more careful numerical treatment \cite{Huang2007PreconditionedDA}. ADMM is a competitive tool in solving problems arising in data science due to its efficiency and computational advantages over other similar methods \cite{ADMM}. In particular, using an ADMM algorithm with a finite element discretized PDE problem can greatly improve practical employability over a more traditional Newton iteration when working on a discrete computational domain represented by a large mesh. 

We now present an ADMM algorithm for the $p$-Poisson problem.  We note that we are minimizing functionals as in \cite{PDEADMM} which is a more general setting than the ADMM setup in \cite{ADMM}. As discussed above, this problem easily becomes ill-conditioned for large $p$-values so it is ideally suited to this approach. For the application of geodesic distance approximation, large $p$-values may be used to yield more accurate results. Thus, it is of interest to use a numerical solver which is robust against ill-conditioned systems.

Fayolle \& Belyaev \cite{FAYOLLE20181} consider $p$-Poisson problems and implement an ADMM algorithm for computing numerical solutions. Our implementation is very similar to that shown by Fayolle \& Belyaev \cite{FAYOLLE20181} but with different boundary conditions. We will see that using ADMM amounts to iterations which require solving a Poisson equation and a 1D polynomial equation.

In order to split the problem, a slack variable $\xi(x)$ is introduced in \cite{FAYOLLE20181}, which gives rise to the constrained problem
\begin{equation*}
    \begin{split}
       & \text{minimize}  \quad \frac1p \int_\Omega |\xi|^p dA - \int_\Omega  u \, dA, \\[3pt]
       & \text{subject to}  \quad  \xi = \gradS u,
    \end{split}
\end{equation*}
with boundary conditions of $u|_{\Gamma_1} = 0$ and $\frac{\p u}{\p n}|_{\Gamma_2} = 0$. Section \ref{sec:model} elaborates on the relation between minimizing the energy functional and these boundary conditions.  In the SFEM discretization, the Dirichlet conditions are imposed by adding constraints to the nodes lying on $\Gamma_1$ and the homogeneous Neumann condition is imposed by disregarding the boundary term obtained from applying the divergence theorem. An ADMM algorithm can now be used on the convex optimization problem. Relaxing the constraint and adding a Lagrange multiplier term gives 
\begin{equation*}
\label{eq:augLagpDist}
\begin{split}
    L_\beta(\xi, u, y) & = \frac1p \int_\Omega |\xi|^p dA - \int_\Omega u dA  +  \int_\Omega y \cdot (\gradS u - \xi ) dA + \frac{\beta}{2} \int_\Omega |\gradS u - \xi |^2 dA \\[3pt]
    & = \int_\Omega \left \{ \frac1p |\xi|^p  -   u  + \frac{\beta}{2}  \left | \xi - \left (\gradS u - \frac{y}{\beta} \right) \right |^2  -\frac{1}{2\beta} |y|^2\right \} dA,
\end{split}
\end{equation*}
where $\beta > 0$ and $y(x)$ is the vector of the Lagrange multipliers. 

First, if $u$ and $y$ are fixed, the optimization with respect to $\xi$ results in a 1D polynomial equation. The optimal $\xi$ will minimize
\begin{equation}
    \frac1p |\xi|^p + \frac{\beta}{2}  \left | \xi - \left (\gradS u - \frac{y}{\beta} \right) \right |^2, 
    \label{eq:xiMin}
\end{equation}
and therefore will be proportional to $\gradS u - \frac{y}{\beta}$, taking the form $c(x) \left( \gradS u(x) -  \frac{y}{\beta} \right)$.
 
Heuristically, minimizing only $\frac1p |\xi|^p$  results in $\xi = 0$ and minimizing only $\frac{\beta}{2}  \left | \xi - \left (\gradS u - \frac{y}{\beta} \right) \right |^2$  results in $\xi =  \gradS u- \frac{y}{\beta}$ so it makes sense that the minimizer will be situated on the straight segment connecting the origin of coordinates with $\gradS u - \frac{y}{\beta} $. To illustrate this, we can separate $\xi$ into parallel and orthogonal components relative to $\gradS u - \frac{y}{\beta} $ denoted, respectively, by $\xi_{\parallel}$ and $\xi_{\perp}$ so that $\xi = \xi_{\parallel}+\xi_{\perp}$. Plugging this into \eqref{eq:xiMin} gives 
\begin{equation*}
 \begin{split}
    \argmin_\xi \left \{ \frac1p |\xi|^p  + \frac{\beta}{2}  \left | \xi - \left (\gradS u - \frac{y}{\beta} \right) \right |^2 \right \} 
    &  \\
    = \argmin_\xi & \left \{ \frac1p |\xi|^p + \frac{\beta}{2}  \left | \xi \right |^2  - \beta \left ( \gradS u - \frac{y}{\beta}  \right )^{\top}  \xi_{\parallel} \right \},
\end{split}
\end{equation*}
from which we can see that the optimal $\xi$ is parallel to $\gradS u - \frac{y}{\beta}$. 

For each $x \in \Omega_h$, using $\xi = c(x) \left( \gradS u(x) -  \frac{y}{\beta} \right)$ leads to minimizing
\begin{equation}
    \frac1p c^p \left |\gradS u - \frac{y}{\beta} \right | ^p + \frac{\beta}{2} \left | \gradS u - \frac{y}{\beta} \right |^2 (c-1)^2, 
    \label{eq:cPreDiff}
\end{equation}
or equivalently solving
\begin{equation}
    \left |\gradS u - \frac{y}{\beta} \right | ^{p-2} c^{p-1}+ \beta (c-1) = 0
    \label{eq:c}
\end{equation}
for $c$ \cite{FAYOLLE20181}. The latter is obtained by differentiating \eqref{eq:cPreDiff} with respect to $c$ and then dividing by $\left | \gradS u - \frac{y}{\beta} \right |^2$. This problem can be dealt with numerically using a Newton iteration. Note that for a given $x \in \Omega_h$, if $|\gradS u(x) - \frac{y}{\beta}| \neq 0$, we will have a root $0 < c < 1$. This is because $c=0$ in \eqref{eq:c} gives $-\beta<0$ and $c=1$ in \eqref{eq:c} gives $|\gradS u- \frac{y}{\beta} |^{p-2} > 0$.

If $\xi$ and $y$ are fixed, optimizing with respect to $u$ leads to solving the Poisson boundary value problem given by
\begin{subequations} \label{eq:pLapu}
\begin{alignat}{3}
        -\Delta^S u = - \Div^S (\xi) - \Div^S \left(\frac{y}{\beta}\right) + \frac1\beta \quad & \text{ in } \Omega,   \label{eq:pLapua}\\
    u = 0 \quad & \text{ on } \Gamma_1,  \label{eq:pLapub}\\
    \frac{\p u}{\p n} = 0 \quad & \text{ on } \Gamma_2, 
    \label{eq:pLapuc}
\end{alignat}
\end{subequations}
which can be done in a standard way using a surface finite element method \cite{FAYOLLE20181}. 

ADMM optimizes the augmented Lagrangian with respect to each variable separately. The ADMM iterative procedure for numerically solving the $p$-Poisson problem is then given by\\

$\bullet$ \underline{$\xi$-update step:} For each $x \in \Omega_h$, solve \eqref{eq:c} with $u=u^k$ and $y=y^k$ to find $c$, then obtain the update $\xi^{k+1} = c(x) \left( \gradS u^k -  \frac{y^k}{\beta} \right)$.\\

$\bullet$ \underline{$u$-update step:} Solve \eqref{eq:pLapu} with $y=y^k$ and $\xi = \xi^{k+1}$ to obtain the update $u^{k+1}$.\\

$\bullet$ \underline{Dual update:} We update the dual variable by $y^{k+1} = y^{k} + \beta (\xi^{k+1}- \gradS u ^{k+1})$. This is because the gradient of $y$ may be computed as  $\gradS y = \xi^*- \gradS u^*$ where star superscripts denote optimal values. Here, we can think of $\beta$ as a step size.  \\

To start the iterative procedure we set $\xi^0 = y ^0= 0$ so $u$ is initialized by solving  $ -\Delta^S u^0 = \frac1\beta$
subject to boundary conditions of $u^0|_{\Gamma_1} = 0$ and $\frac{\p u^0}{\p n}|_{\Gamma_2} = 0$.  \\

Under certain convexity and existence assumptions given by Gabay and Mercier \cite{PDEADMM}, convergence may be established. Glowinski and Marrocco \cite{Glowinski1974OnTS} verify that the $p$-Poisson problem satisfies necessary assumptions on the plane. 
Provided these assumptions hold, the ADMM iterates satisfy the following: (i) $r^k = \xi^k-\gradS u^k \to 0$ as $k \to \infty$, i.e., the iterates approach feasibility, (ii) the objective function of the iterates approaches the optimal value as $k \to \infty$, and (iii) $y^k \to y^*$ as $k \to \infty$, where $y^*$ is a dual optimal point \cite{ADMM}. Our numerical simulations exhibit convergence of the algorithm.

In addition to tracking the change in the computed solution between iterates, the primal and dual residuals can help inform stopping criteria. The `primal residual' $r^k$ is given by 
\begin{equation*}
    r^k = \xi^k-\gradS u^k = (y^k- y^{k-1})/\beta,
\end{equation*}
and the `dual residual' $s^k$ is given by 
\begin{equation*}
    s^k = -\beta (\xi^k- \xi^{k-1}).
\end{equation*}
It is often reasonable to use a stopping condition of 
$||r^k|| < tol_{primal} \text{ and } ||s^k|| < tol_{dual}$ where $tol_{primal}$ and $tol_{dual}$ are chosen by the user. For most test problems, we set $tol_{dual} = 10^{-3}$ and $tol_{primal} = 10^{-6}$. Note that a smaller penalty parameter, $\beta$, often yields a better dual residual and a larger $\beta$ often yields a better primal residual. We wish to strike a healthy balance; typically, we use $\beta = 10$.


\section{Computational assessment} 
\label{sect:numerics}

This section numerically assesses the ADMM algorithm for the $p$-Poisson distance-to-feature problem. The algorithm is implemented in C++ using deal.II \cite{dealII94}. Visualizations shown are performed with ParaView \cite{paraview}. For smooth examples, we can anticipate numerical convergence of order $\mathcal{O}\left(h^2 + \frac1p \right)$ where $h$ is the maximum cell edge length. We use $\ell$ to denote the average cell edge length on the mesh. However, distance functions are generally nonsmooth leading us to observe some deterioration in the convergence rates. In particular, distance functions will be nonsmooth near $\Gamma_1$, e.g., distance to the origin in 1D gives the absolute value function.  

We numerically compute the distance-to-feature for several examples, using three methods: our $p$-Poisson method (using a low and a large value of $p$, respectively), the \heatmethod from Crane at al. \cite{Crane}, and the classical polyhedral method from Mitchell et al. \cite{Mitchell}. We use the C++ implementations of these methods from geometry-central \cite{geometrycentral} and export the results to ParaView \cite{paraview} for consistent visualization. We use the same surface triangulation for all methods.
 
We emphasize that the $p$-value can control the smoothness of our $p$-Poisson distance approximations. This allows us to obtain smoothed distance function representations which may be desired in some practical applications \cite{Crane}.  Additionally, by taking a larger $p$-value, we recover the nonsmooth nature of the true distance function. For a visualization of small and large $p$-values shown side-by-side, see Figures~\ref{fig:ExactExamples} and \ref{fig:manyEx}. 

\subsection{Numerical convergence studies} 
\label{sec:convstudies}
In this subsection we present numerical convergence studies for some examples on simple surfaces where we know the exact distance functions, $\mathrm{dist}(\cdot, \Gamma_1)$, to the feature sets. 

The symmetric mean absolute percentage error (SMAPE) \cite{Nguyen2019}  is given by
\begin{equation*}
    \text{SMAPE}(\mathrm{dist}, u_p) = \frac{100}{n} \sum_{i=1}^n \frac{|\mathrm{dist}^i - u_p^i|}{\left( |\mathrm{dist}^i| + |u_p^i|\right)/2} , 
\end{equation*}
where $u_p^i$ is the numerically computed $p$-Poisson distance at vertex $i$, and $\mathrm{dist}^i$ denotes the exact value of the geodesic distance from the vertex $i$ to the feature set. Here, $n$ denotes the number of vertices.

For a well-rounded error assessment, we also present $L^2$ integral norm relative errors between the computed $p$-Poisson distance, $u_p$, and the exact geodesic distance function, $\mathrm{dist}(\cdot, \Gamma_1)$. 

\subsubsection{Distance to a point on the hemisphere}
\label{subsect:hemiPT}
The first example we consider is the geodesic distance to a point on the unit hemisphere. In this example, $S = \{(x,y,z) : x^2+y^2+z^2 =1, x>0\}$ is the unit hemisphere, 
$\Gamma_1 = \left \{ \left( \frac{\sqrt{2}}{2}, \frac12, \frac12 \right) \right \} $, and $\Gamma_2 = \p S$. The geodesic distance between two points on a sphere is given by the spherical law of cosines. In particular, the geodesic distance between a point $\mathbf{x} = (x,y,z)$ on the hemisphere and $\Gamma_1$ is given by 
\begin{equation}
    \textrm{dist}\left(\mathbf{x},\Gamma_1\right) = \arccos{\left( \frac{\sqrt{2}}{2} x + \frac12 y + \frac12 z \right)}.
    \label{eq:exPtHemi}
\end{equation}

\begin{figure}
\centering           
\begin{multicols}{2}
    $p = 5$ \hspace{6em} $p = 100$ \\
    \hrulefill 
    
    \includegraphics[width=0.47\linewidth]{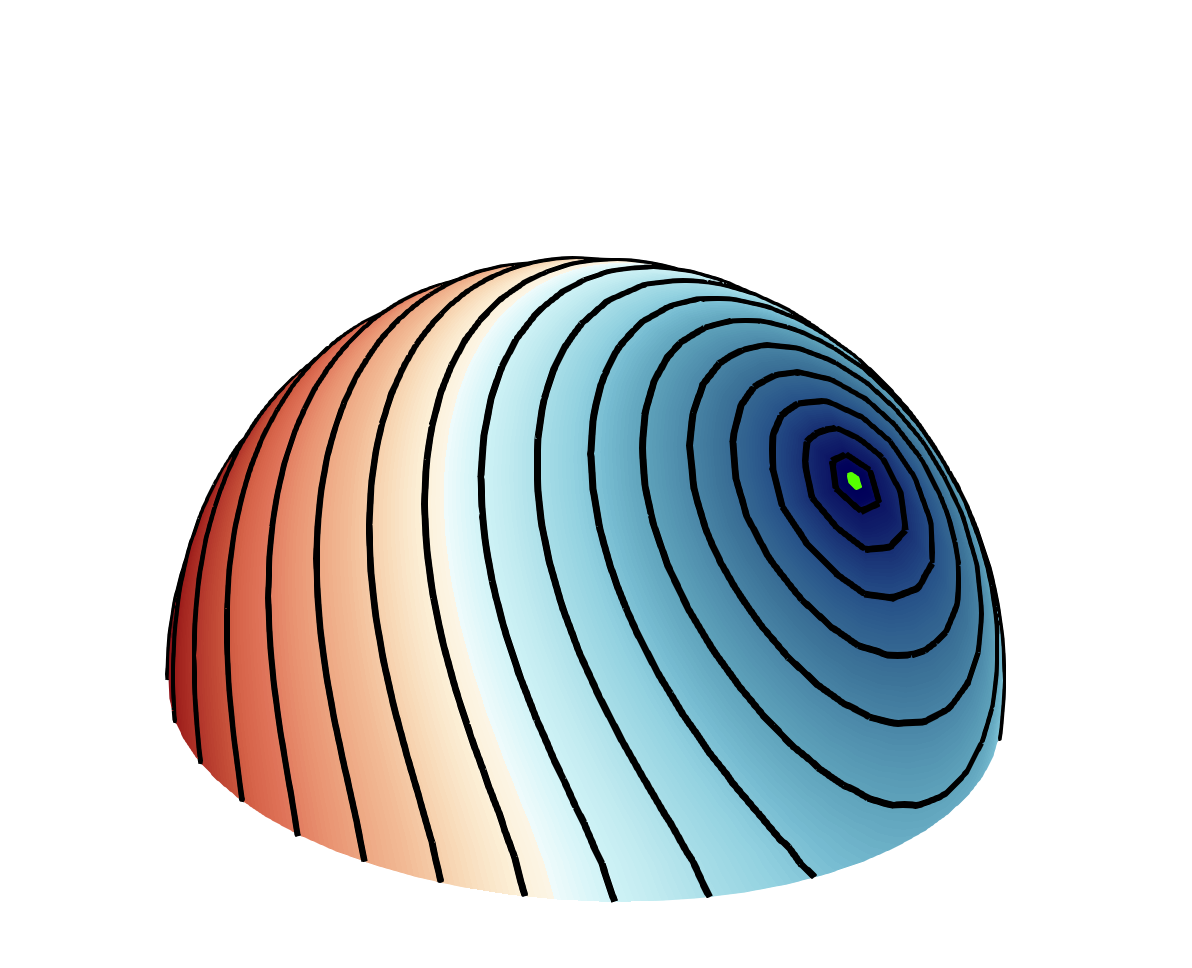}
    \includegraphics[width=0.47\linewidth]{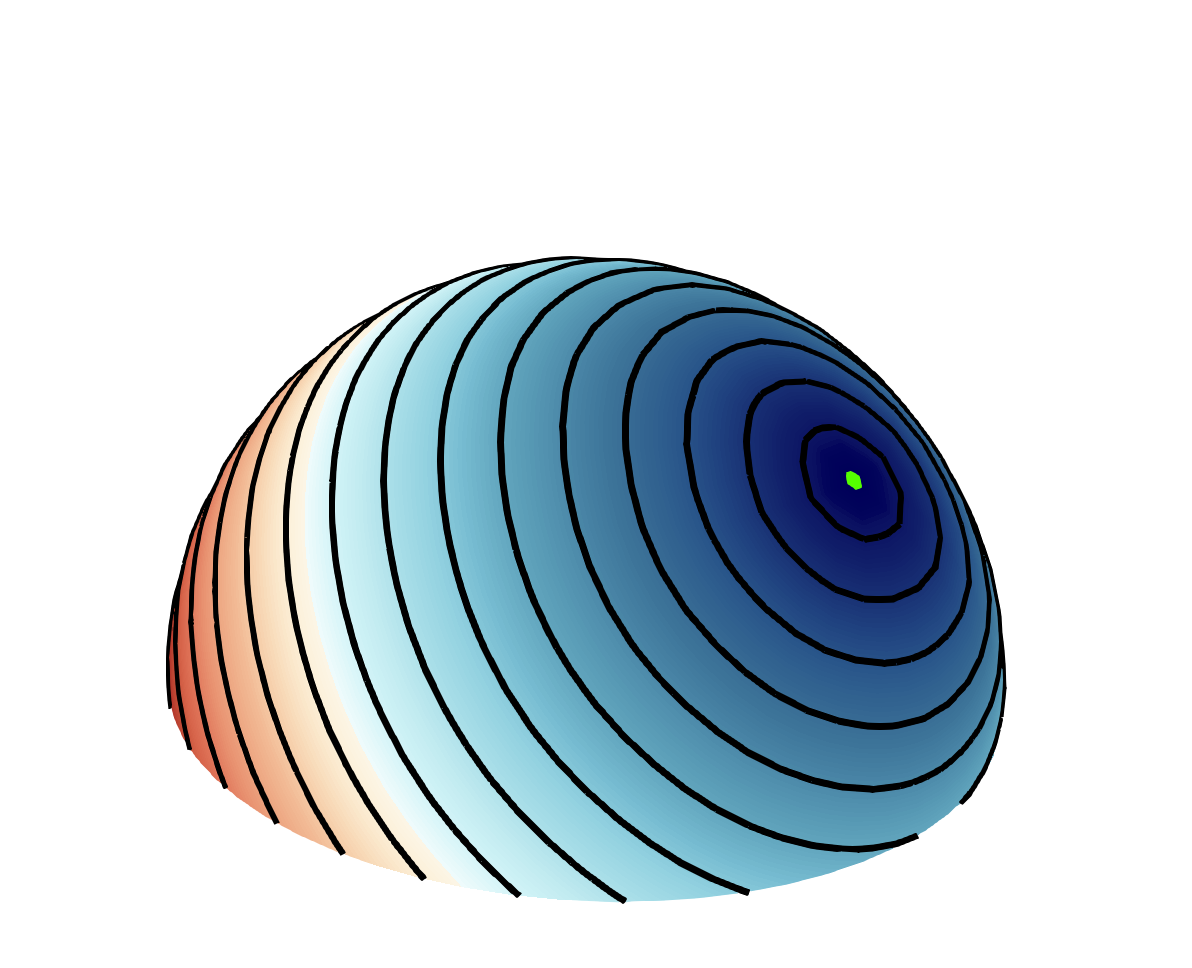}

    \includegraphics[width=0.4\linewidth]{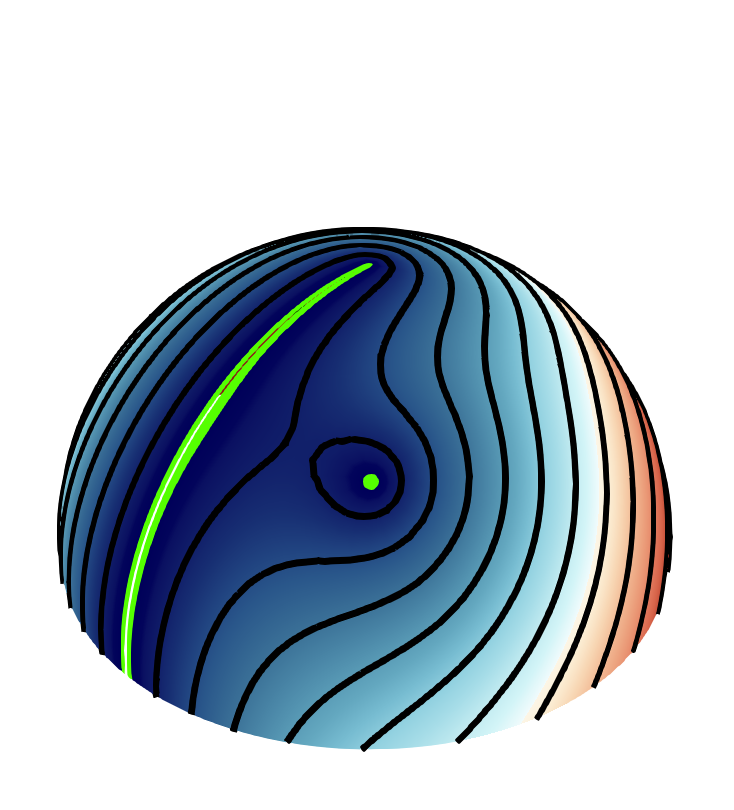}
    \hspace{1em}
    \includegraphics[width=0.4\linewidth]{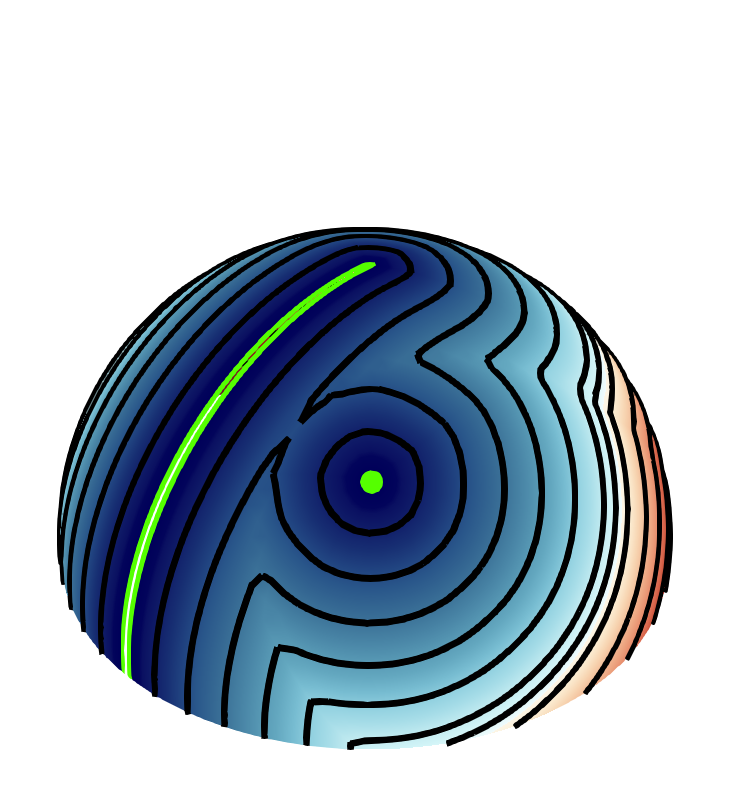}

    \includegraphics[width=0.4\linewidth]{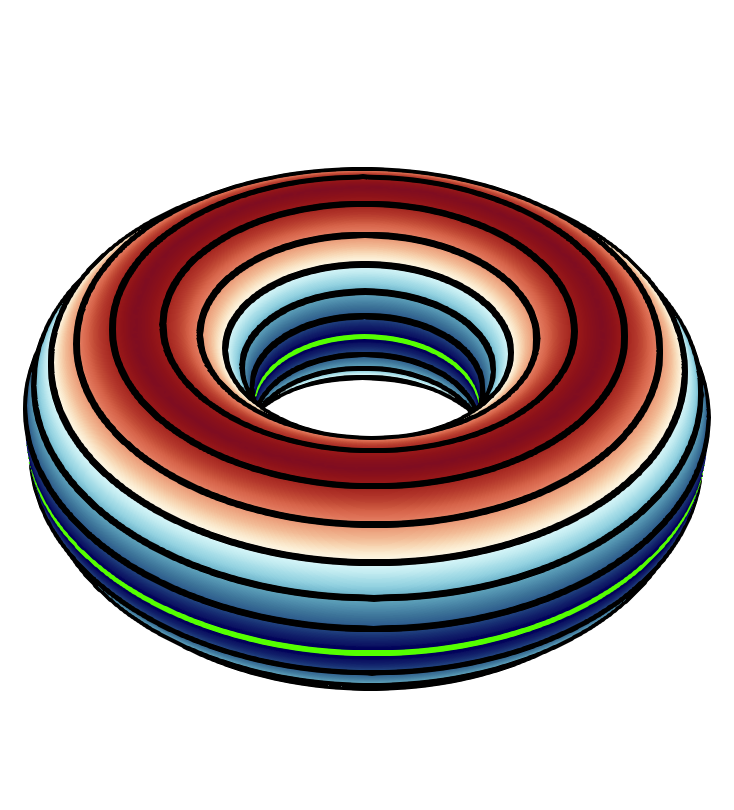}
    \hspace{1em}
    \includegraphics[width=0.4\linewidth]{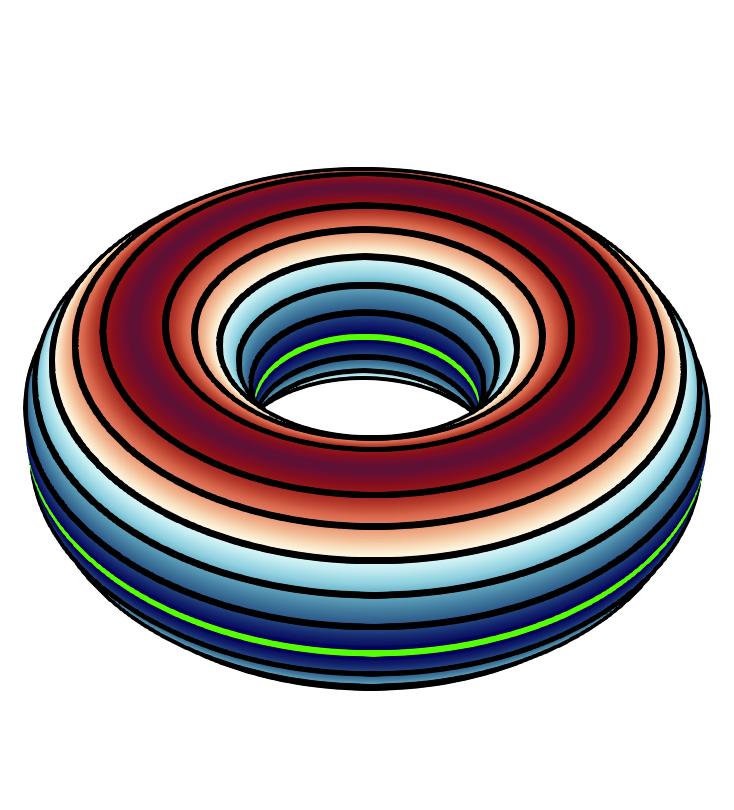}
    \columnbreak \\
    heat distance \cite{Crane} \hspace{2em} polyhedral \cite{Mitchell} \\
    \hrulefill 
    
   \includegraphics[width=0.47\linewidth]{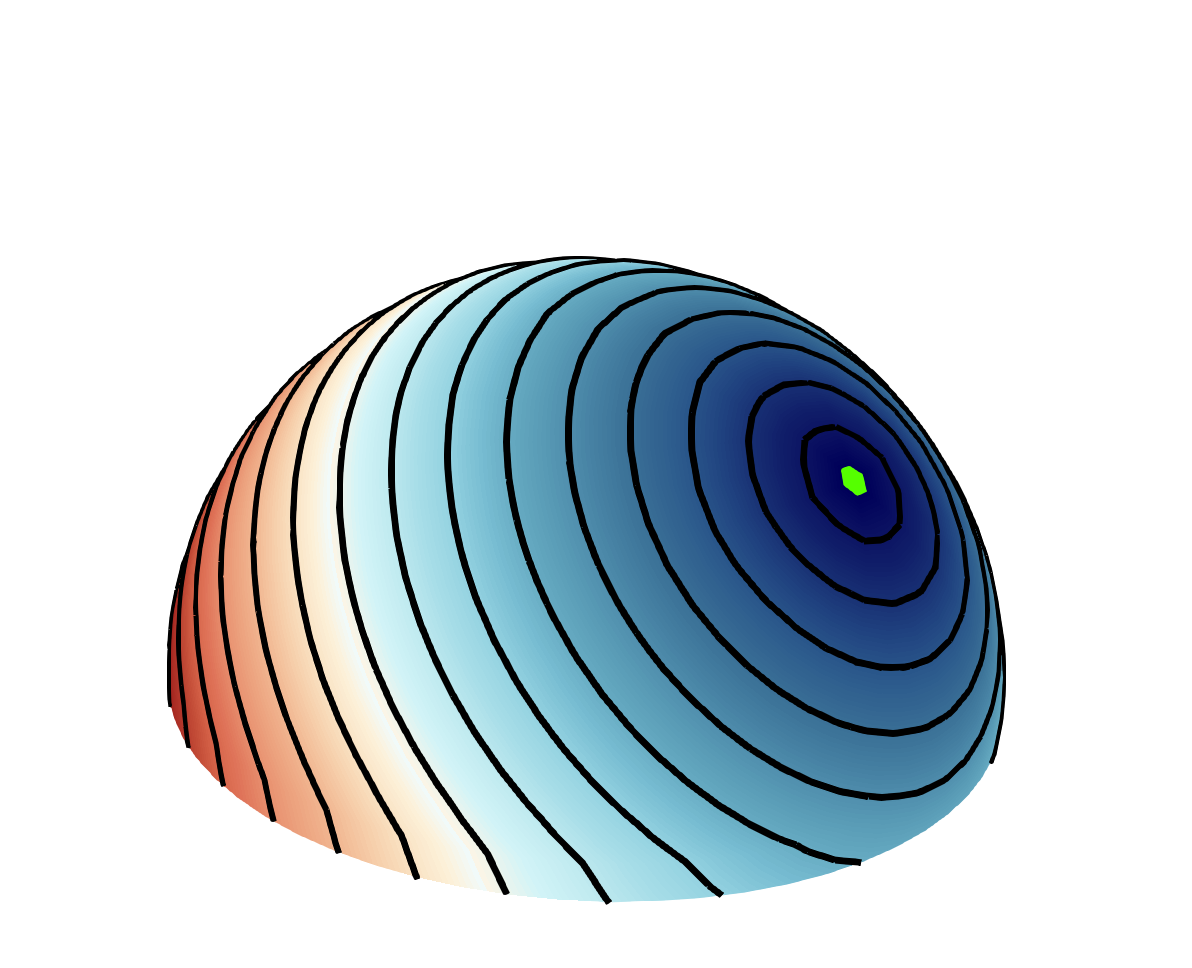}
    \includegraphics[width=0.47\linewidth]{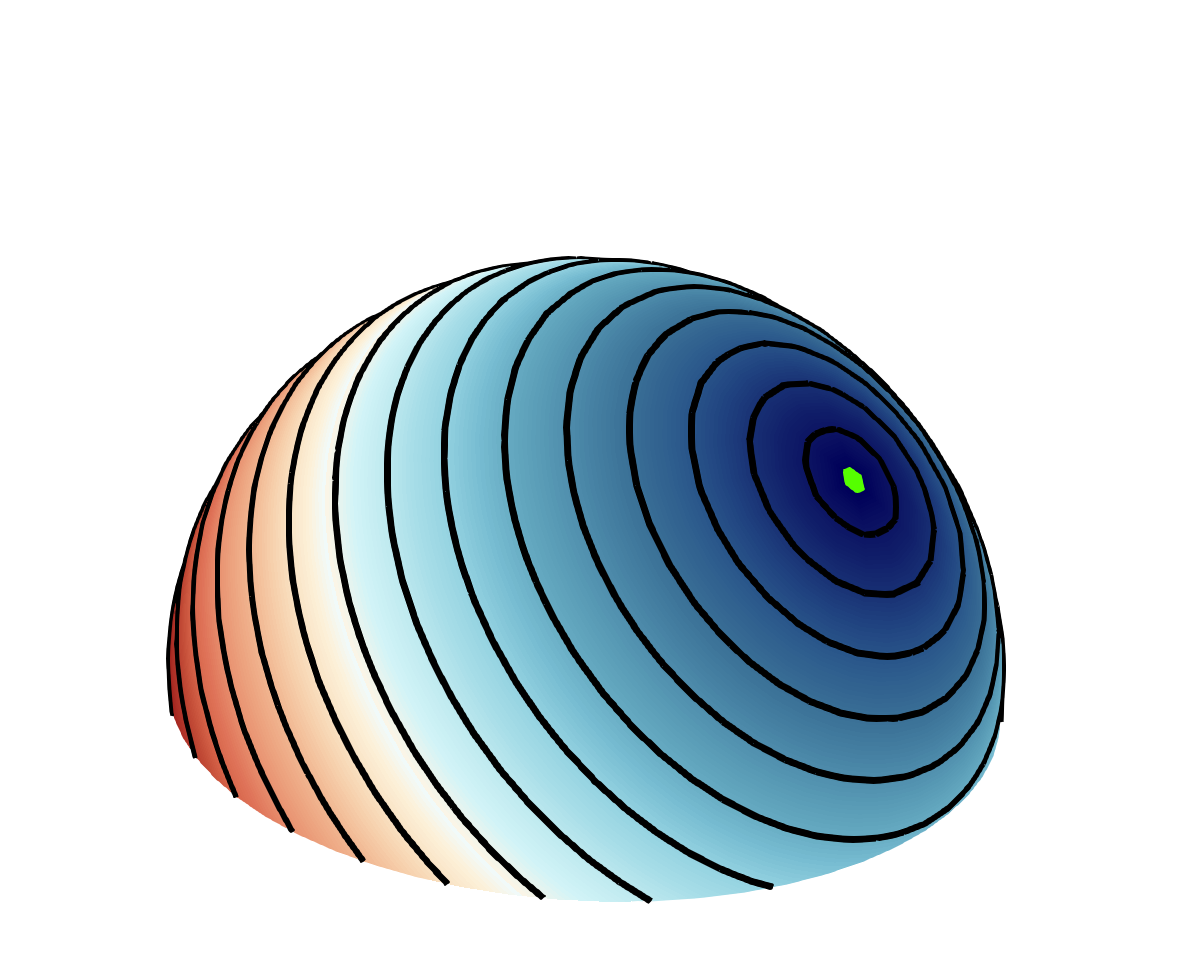}

    \includegraphics[width=0.4\linewidth]{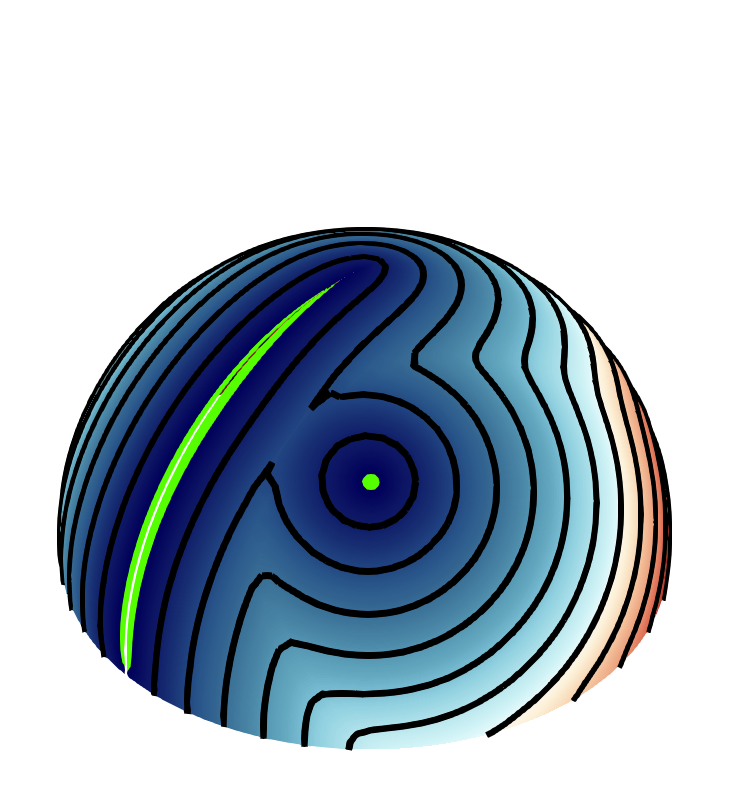}
    \hspace{1em}
    \includegraphics[width=0.4\linewidth]{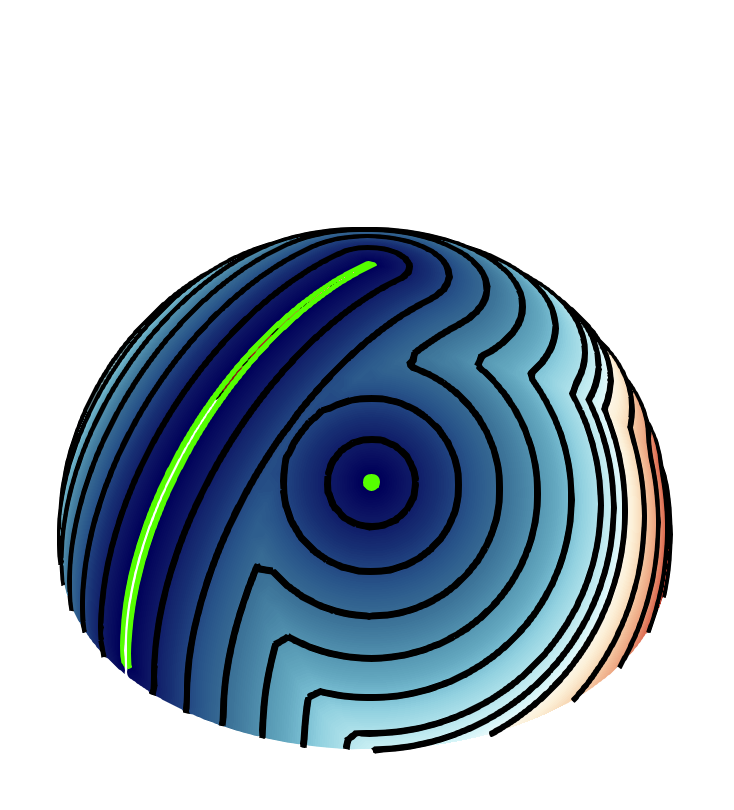}

    \includegraphics[width=0.4\linewidth]{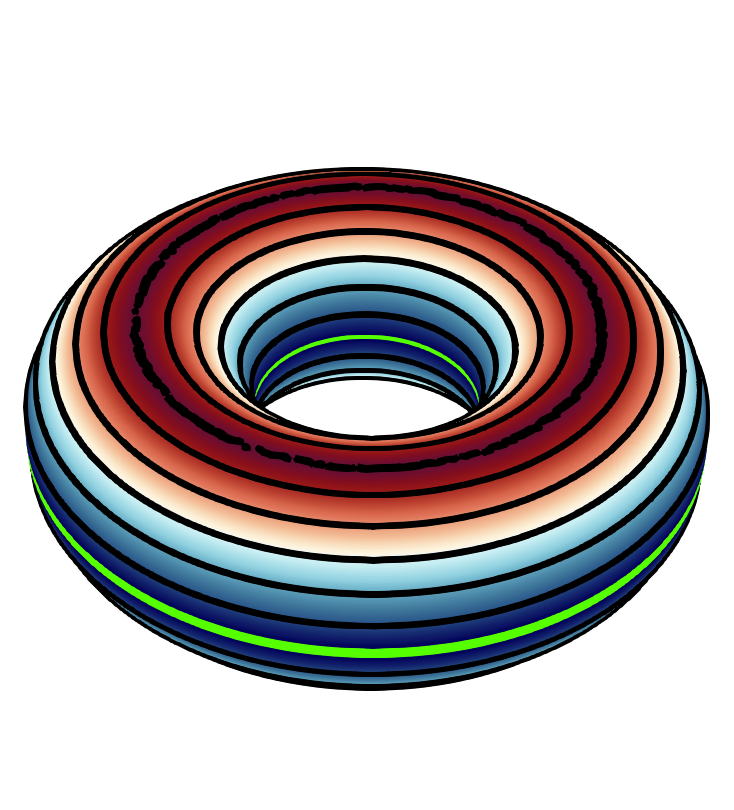}
    \hspace{1em}
    \includegraphics[width=0.4\linewidth]{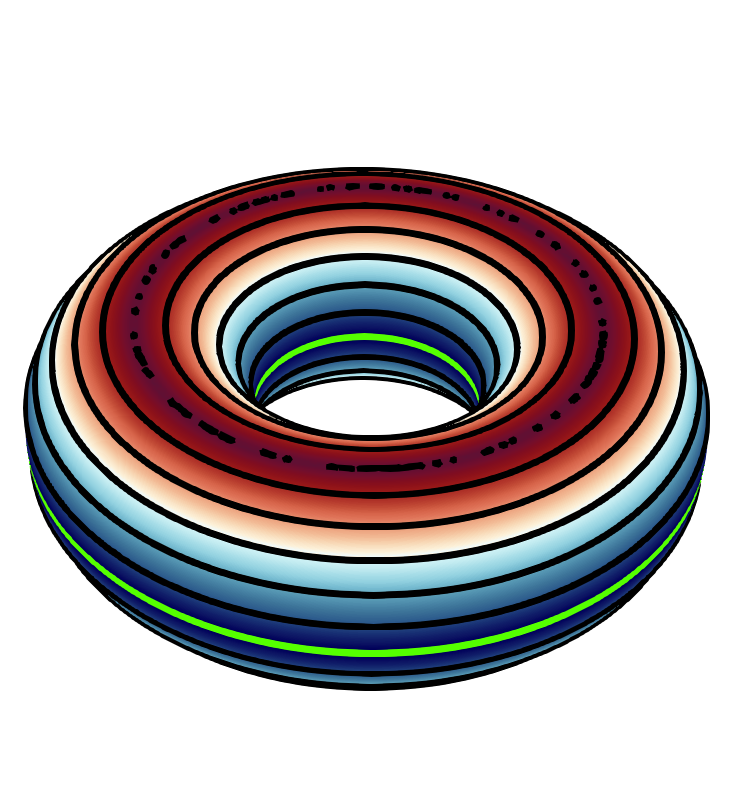}
    \end{multicols}

    \caption{Computed distance approximations using three methods: our $p$-Poisson method (first and second columns, using $p=5$ and $p=100$, respectively), the \heatmethod (third column), and the polyhedral method (fourth column). Top row: distance to a point on hemisphere. Second row: distance to an open curve and a point on hemisphere. Bottom row: distance to two closed curves on torus. For visualization near the feature set $\Gamma_1$, we highlight in green values falling below a chosen tolerance. The triangulations consists of $2307$ vertices and $791$ faces, $9484$ vertices and $3228$ faces, and $35019$ vertices and $105057$ faces respectively. }
    \label{fig:ExactExamples}
\end{figure}

The first row of \Cref{fig:ExactExamples} shows the computed distance-to-feature in this example, by the three methods. The order (from left to right) is as follows: the first and second plots correspond to the $p$-Poisson distance approximation using $p=5$ and $p=100$, respectively; the third plot is for the heat method, and the fourth shows the polyhedral distance \cite{Mitchell}. For the $p$-Poisson method with $p = 5$, the contours are less evenly spaced and the Neumann condition on $\Gamma_2$ is visually prominent. For $p = 100$, the Neumann condition is visually imperceptible and the contours are evenly spaced. For the \heatmethod we note a severe effect of the boundary on the contours. The polyhedral method seems to give comparable results to the $p$-Poisson method with $p=100$.

\begin{figure}
    \centering
    \includegraphics[scale = 0.6]{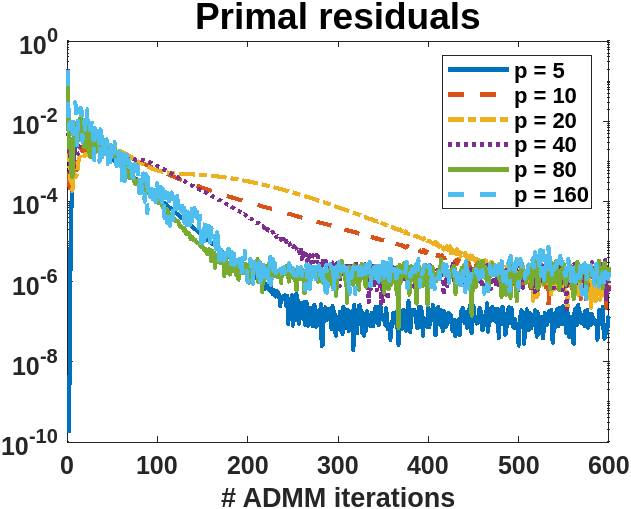} \qquad 
    \includegraphics[scale = 0.6]{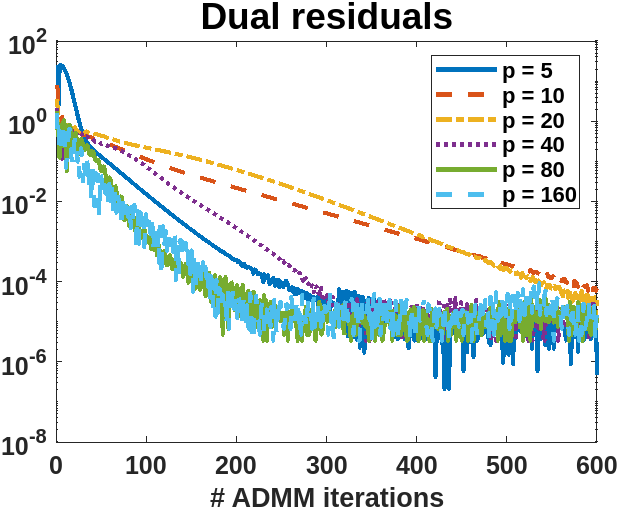}\\
    \caption{Tracking of residuals and gradient norm for the example of distance to a point on the hemisphere using various $p$-values. Primal (left) and dual (right) residuals versus ADMM iterations. The algorithm is run on a hemisphere mesh with $10240$ tri cells and $\ell \approx 3.95 \times 10^{-2}$, $h \approx 6.29 \times 10^{-2}$.  
    }
    \label{fig:resSPHPT}
\end{figure}

\Cref{fig:resSPHPT} shows how the primal and dual residuals evolve with the number of ADMM iterations for various $p$-values. For each $p>5$, we initialize with the computed solution from the previous (smaller) $p$-value shown in \Cref{fig:resSPHPT}. Since we have already computed the solution for the previous (smaller) $p$-value, this initialization process reduces the number of iterations needed for the current (larger) $p$-value. However, this is not necessary to see convergence using the ADMM scheme. Unlike a Newton algorithm, ADMM will converge even with a poor initial guess. 

\subsubsection{Distance to an open curve and a point on the hemisphere}
\label{subsect:curve-point}
We now consider an example that seeks the distance to $\Gamma_1$ consisting of an open curve and a point on the unit hemisphere. Specifically,  $S = \{(x,y,z) : x^2+y^2+z^2 =1, x>0\}$ is the hemisphere, 
\[
\Gamma_1 = \left \{ \left( \sqrt{2}/{2}, 1/2, 1/2 \right) \right \} \cup \left \{ (x, y, z) \in S: z = 0 , y \geq 0 \right \}, 
\]
and $\Gamma_2 = \p S \setminus \left \{ \left( 0, 1, 0 \right) \right \}$ -- see the second row in \Cref{fig:ExactExamples}. 

The second row of \Cref{fig:ExactExamples} shows the numerical computations for this example, using the three methods (the order of the plots is similar to the first row). For the $p$-Poisson method, we observe that the contours sharpen for larger $p$-values and the influence of the Neumann condition diminishes. In particular, for $p=5$, we see smoothing around points that are equidistant to the point and the open curve which comprise $\Gamma_1$. The smoothing of the contours is also clearly visible in the heat method. However, for $p = 100$, there is no perceptible smoothing of the distance, and a similar comment applies to the polyhedral method as well. 

\Cref{tab:hemiCOMPopenPT} shows $L^2$ relative errors and SMAPE for our method applied to the problem of computing geodesic distance to an
 open curve and a point on the hemisphere. Here we use an analytical expression for the geodesic distance, derived from the spherical law of cosines, to compare against our computed solutions. Our primary focus is on the convergence of our method with respect to $p$.  Thus, computations are performed on a large, fixed number of cells, iterating until convergence of
the ADMM algorithm.   Computations are performed by successively doubling $p$ at each iteration.  The results show linear convergence in $1/p$ for both error metrics.  
It is worth noting that the SMAPE table entries are roughly 100 times larger than the $L^2$ entries because SMAPE represents a percentage error.

\begin{table}
\begin{center}
\begin{tabular}{|c|c|c|c|c|}
    \hline 
     \multicolumn{5}{|c|}{ADMM algorithm for $p$-Poisson distance on hemisphere}\\
     \hline
     $p$ &  $L^2$ (relative error) & Rate  & SMAPE (\% error)  & Rate \\
     \hline
     $5$ & $1.2928 \times 10 ^{-1}$ & $NA$ & $16.301$ & $NA$  \\
     \hline 
     $10$ &  $6.2872 \times 10 ^{-2}$ & $1.04$ & $8.1910$ & $0.99$ \\
     \hline 
     $20$ &  $3.0476 \times 10 ^{-2}$ & $1.04$ & $3.9934$ & $1.04$   \\
     \hline 
     $40$ &  $1.4554 \times 10 ^{-2}$ & $1.07$ & $1.9136$ & $1.06$ \\
     \hline 
     $80$ &  $6.9663 \times 10 ^{-3}$ & $1.06$ & $9.6856 \times 10 ^{-1}$ & $0.98$ \\
     \hline
     $160$ &  $3.4894 \times 10 ^{-3}$ & $1.00$ & $4.8681 \times 10 ^{-1}$ & $0.99$ \\
     \hline
\end{tabular}

\end{center}
    \caption{Numerical convergence study in $p$, comparing computed solutions with exact geodesic distance to an open curve and a point on the hemisphere. The algorithm is run on a hemisphere mesh with $655360$ tri cells and $\ell \approx 9.89 \times 10^{-3}$, $h \approx 1.57 \times 10^{-2}$. We note that the SMAPE values are \% errors and are thus about 100 times larger than the $L^2$ relative error values. }
    \label{tab:hemiCOMPopenPT}
\end{table}

\subsubsection{Distance to closed curves on torus}
\label{subsect:tor}

This example seeks the distance to $\Gamma_1$ consisting of two closed curves on a torus. Here, $S = \{(x,y,z) : ( R - \sqrt{x^2 + z^2} )^2 + y^2 = r^2 \}$ is the torus with  outer radius $R=2$ and inner radius $r=1$, $\Gamma_1$ consists of the inner and outer circles on the torus lying in the plane $y = 0$, and $\Gamma_2 = \emptyset $ -- see the bottom row in \Cref{fig:ExactExamples}. 

The bottom row of \Cref{fig:ExactExamples} showcases the numerical computations for this example (same three methods and same layout of plots as for the first two rows). Note that since the torus is a closed surface, there is no error from the Neumann boundary condition for the $p$-Poisson method. The lower $p$-value gives a smoother approximation, and the higher value gives a sharper, more accurate approximation. The heat and polyhedral methods give similar results to the $p$-Poisson method with $p=100$. 

\Cref{tab:tor} shows $L^2$ relative errors and SMAPE for our method applied to computing the geodesic distance to closed curves defined by $y=0$ on a torus.  An analytical expression for the geodesic distance is used to compute the errors. As in the previous example, computations are performed on a large, fixed number of cells, iterating until convergence of the ADMM algorithm. For each experiment, $p$ is successively doubled.
Once again, we find linear convergence in $1/p$ for both error metrics. 

\begin{table}
\begin{center}
\begin{tabular}{|c|c|c|c|c|}
    \hline 
     \multicolumn{5}{|c|}{ADMM algorithm for $p$-Poisson distance on torus}\\
     \hline
     $p$ &  $L^2$ (relative error) & Rate  & SMAPE (\% error)  & Rate \\
     \hline
     $5$ & $5.8144 \times 10 ^{-2}$ & $NA$ & $ 7.0674$ & $NA$  \\
     \hline 
     $10$ &  $2.9089 \times 10 ^{-2}$ & $1.00$ & $3.3470$ & $1.08$ \\
     \hline 
      $20$ &  $1.4132 \times 10 ^{-2}$ & $1.04$ & $1.6215$ & $1.05$   \\
     \hline 
    $40$ &  $6.4439 \times 10 ^{-3}$ & $1.13$ & $7.7920 \times 10 ^{-1}$ & $1.06$ \\
     \hline 
    $80$ &  $2.9304 \times 10 ^{-3}$ & $1.14$ & $3.7879 \times 10 ^{-1}$ & $1.04$ \\
     \hline
      $160$ &  $1.5206 \times 10 ^{-3}$ & $0.95$ & $1.9328 \times 10 ^{-1}$ & $0.97$ \\
     \hline
\end{tabular}

\end{center}
    \caption{Numerical convergence study in $p$, comparing computed solutions with exact geodesic distance to closed curves on a torus. The algorithm is run on a torus mesh with $196608$ tri cells and $\ell \approx 1.65 \times 10^{-2}$, $h \approx 2.74 \times 10^{-2}$. We note that the SMAPE values are \% errors and are thus about 100 times larger than the $L^2$ relative error values. }
    \label{tab:tor}
\end{table}

\begin{figure}
    \centering
    \includegraphics[scale = 0.9]{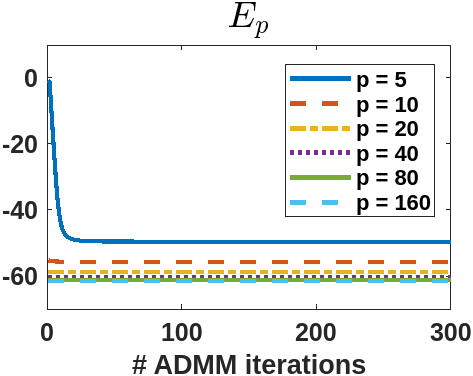}
    \caption{$p$-Poisson energy \eqref{eq:Ep} versus ADMM iterations for distance to closed curves on torus using various $p$-values. The algorithm is run on a torus mesh with $196608$ tri cells and $\ell \approx 3.30 \times 10^{-2}$, $h \approx 5.49 \times 10 ^{-2}$. 
    }
    \label{fig:EpTOR}
\end{figure}

We may also monitor how the $p$-Poisson energy \eqref{eq:Ep} evolves with the ADMM iterations. \Cref{fig:EpTOR} plots the energy versus the number of iterations for a variety of $p$-values. As with the residuals, computations in \Cref{fig:EpTOR} are initialized for each $p>5$ with the computed solution from the previous (smaller) $p$-value. We see that the energy decays rapidly within the first few iterations and then remains level at the (supposed) minimizer.  Other examples exhibit similar results. If the energy begins to increase, it is often a sign that the $p$-value is too large for the mesh resolution or that the choice of parameters is unsuitable; for instance, taking the penalty parameter $\beta$ too small.  

Our numerical simulations were performed on a single thread of an AMD Ryzen 7 8845HS (16-thread) Lenovo Yoga Pro 7 14AHP9. For this torus example, an average ADMM iteration took about $0.049$ seconds on a mesh with $1536$ vertices and $4608$ faces, and about $0.185$ seconds on a mesh with $6144$ vertices and $18432$ faces. We observed that the computational timing remains stable as the $p$-value increases. For successive refinements, the computation time increased by approximately a factor of $4$.

Table \ref{tab:method-stats} shows side-by-side statistics for our $p$-Poisson method, the heat method, and the polyhedral method. For comparison, we fix a large value of $p = 160$. Both $L^2$ relative errors and SMAPEs are measured against the exact geodesic distance function. For Example \ref{subsect:hemiPT}, the geodesic distance function is given by the spherical law of cosines (see (\ref{eq:exPtHemi})). Similarly, we have explicit geodesic distance functions for Example \ref{subsect:curve-point} and Example \ref{subsect:tor}. Table \ref{tab:method-stats} shows that in all three examples, our $p$-Poisson method yields errors that lie between those of the heat method and the polyhedral method. For Example \ref{subsect:curve-point} and Example \ref{subsect:tor}, the SMAPE produced by the heat method is comparable to that of the $p$-Poisson method with $p \in (10, 20)$ and $p \in (40, 80)$ respectively, as shown in Table \ref{tab:hemiCOMPopenPT} and Table \ref{tab:tor}.

\begin{table}[]
    \centering
    \begin{tabular}{|c|c|c|}
    \hline 
     \multicolumn{3}{|c|}{Example \ref{subsect:hemiPT}:  Distance to point on the hemisphere }\\
     \hline
      Method &  $L^2$ (relative error) & SMAPE (\% error) \\
       \hline 
       $p$-Poisson ($p =  160$) & $4.7582 \times 10^{-3}$ &  $7.0265 \times 10^{-1}$\\
       \hline 
       heat distance \cite{Crane} &  $1.0743 \times 10^{-2}$ &  $1.0852$\\
       \hline 
       polyhedral \cite{Mitchell} & $8.7917 \times 10^{-5}$ & $2.3249 \times 10^{-2}$ \\ 
       \hline
    \hline 
     \multicolumn{3}{|c|}{Example \ref{subsect:curve-point}:  Distance to an open curve and a point on the hemisphere }\\
     \hline
      Method &  $L^2$ (relative error) & SMAPE (\% error) \\
       \hline 
       $p$-Poisson ($p =  160$) & $3.4894 \times 10 ^{-3}$ &  $4.8681 \times 10 ^{-1}$ \\
       \hline
        heat distance \cite{Crane} &  $2.7041 \times 10^{-2}$ &  $4.8715$\\
       \hline 
       polyhedral \cite{Mitchell} & $ 2.0248 \times 10^{-4}$ & $3.3905 \times 10^{-2}$ \\ 
       \hline
    \hline 
     \multicolumn{3}{|c|}{Example \ref{subsect:tor}:  Distance to closed curves on torus}\\
     \hline
      Method &  $L^2$ (relative error) & SMAPE (\% error) \\
       \hline 
       $p$-Poisson ($p =  160$) &   $1.5206 \times 10 ^{-3}$ & $1.9328 \times 10 ^{-1}$ \\
       \hline 
       heat distance \cite{Crane} &  $2.4888 \times 10^{-3}$ &  $6.6466 \times 10^{-1}$\\
       \hline 
       polyhedral \cite{Mitchell} & $6.6534 \times 10^{-4}$ & $ 1.2683 \times 10^{-2}$ \\ 
       \hline
    \end{tabular}
    \caption{Side-by-side method statistics ($L^2$ relative error and SMAPE) comparing against the exact geodesic distance function; see, for example, (\ref{eq:exPtHemi}). The hemisphere mesh for Examples \ref{subsect:hemiPT}-\ref{subsect:curve-point} consists of $655360$ tri cells, with $\ell \approx 9.89 \times 10^{-3}$, $h \approx 1.57 \times 10^{-2}$. The torus mesh for Example \ref{subsect:tor} consists of $196608$ tri cells, with $\ell \approx 1.65 \times 10^{-2}$, $h \approx 2.74 \times 10^{-2}$. }
    \label{tab:method-stats}
\end{table}

\subsection{Examples on various surfaces}
We now show examples on a wider variety of surfaces. Additionally, we computationally assess robustness to poor mesh quality and check that our distance approximations satisfy the triangle inequality. 

\begin{figure}
    \centering

\begin{multicols}{2}
    $p = 5$ \hspace{6.5em} $p = 100$ \\
    \hrulefill 

    \includegraphics[width=0.45\linewidth]{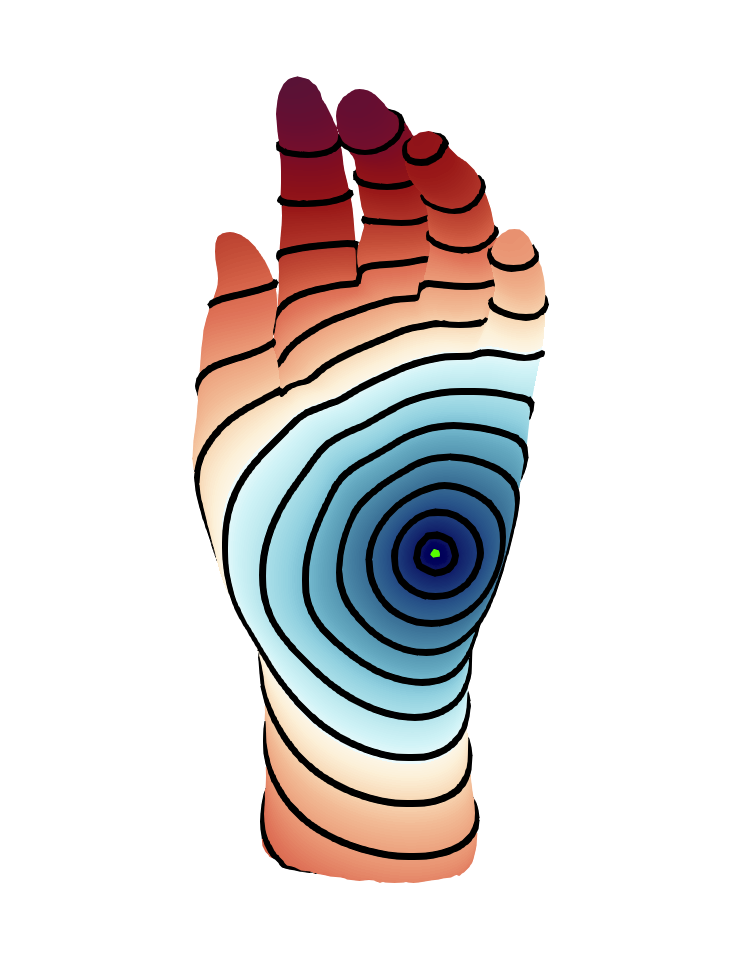}
     \includegraphics[width=0.45\linewidth]{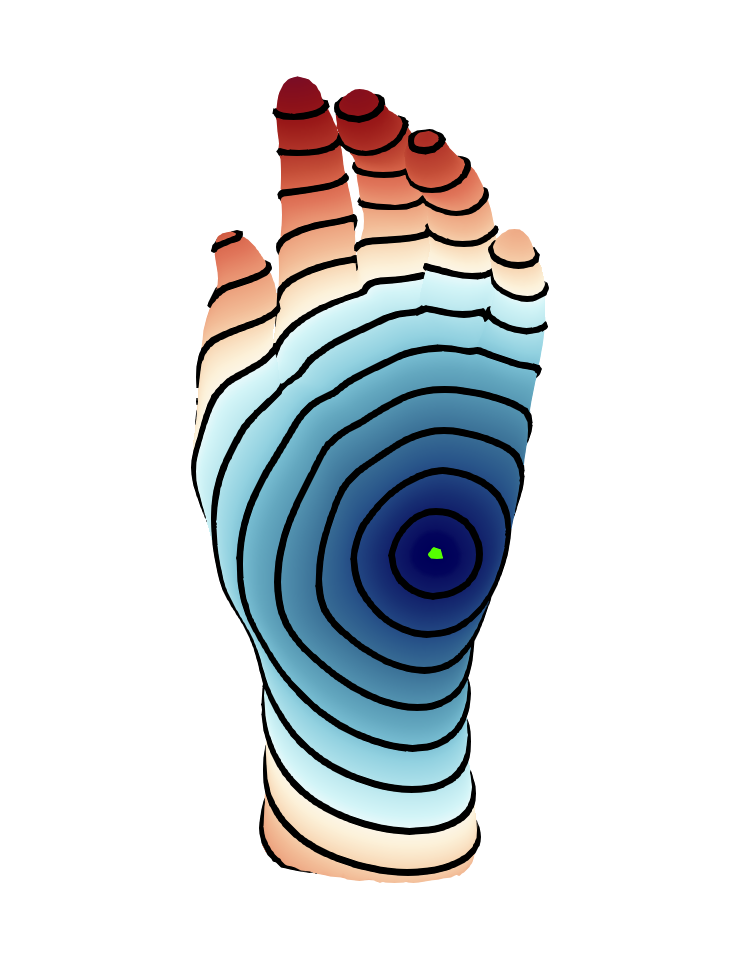}\\

    \includegraphics[width=0.49\linewidth]{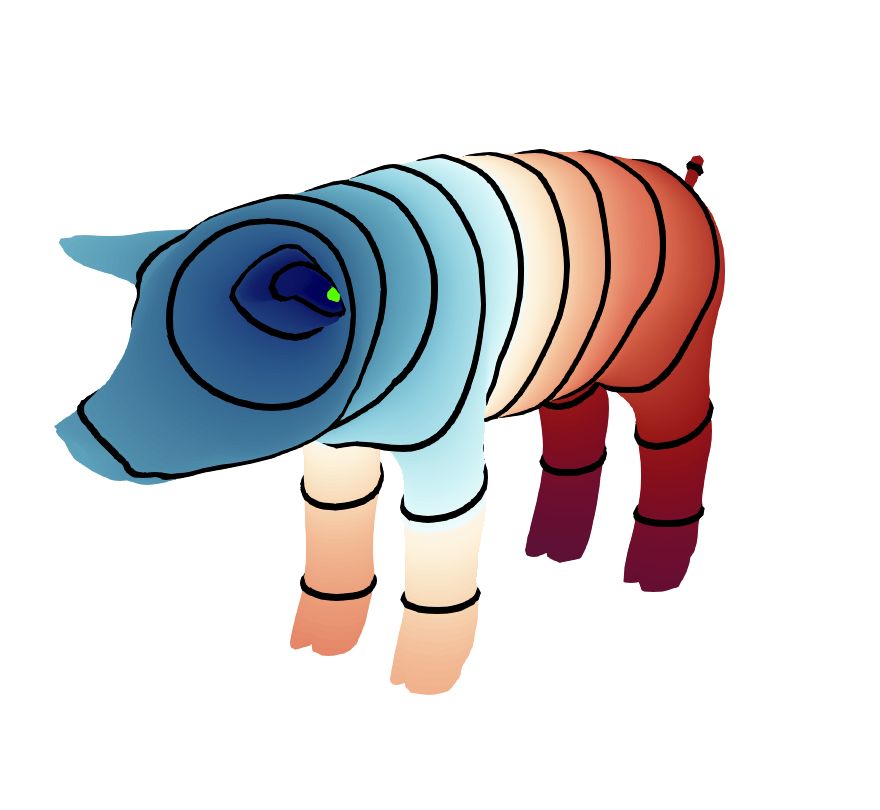}
    \includegraphics[width=0.49\linewidth]{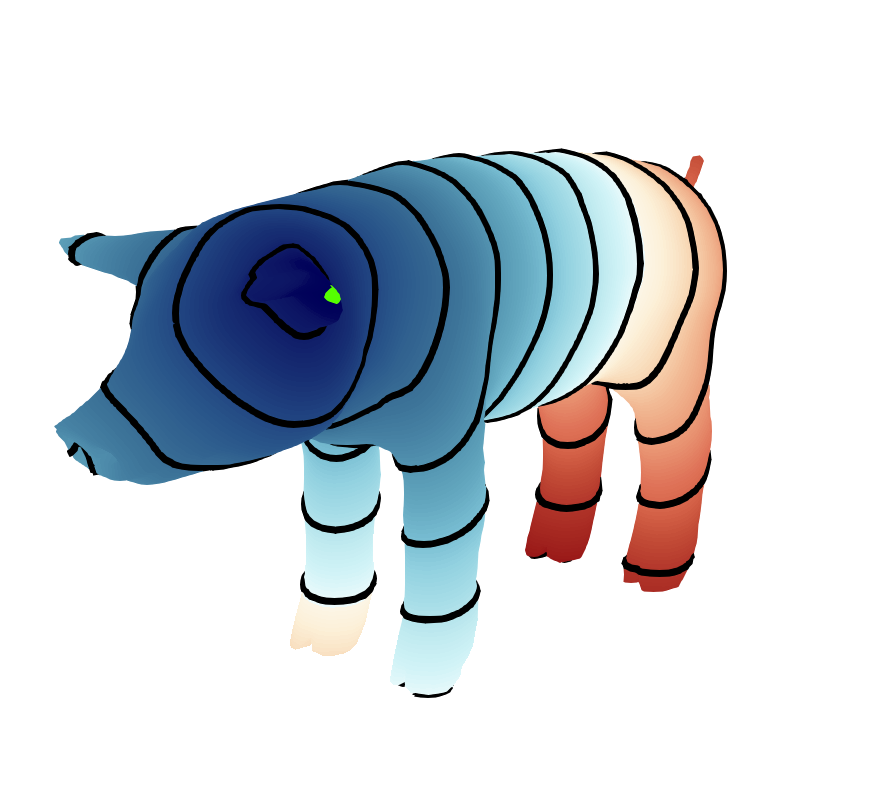}\\

    \includegraphics[width=0.49\linewidth]{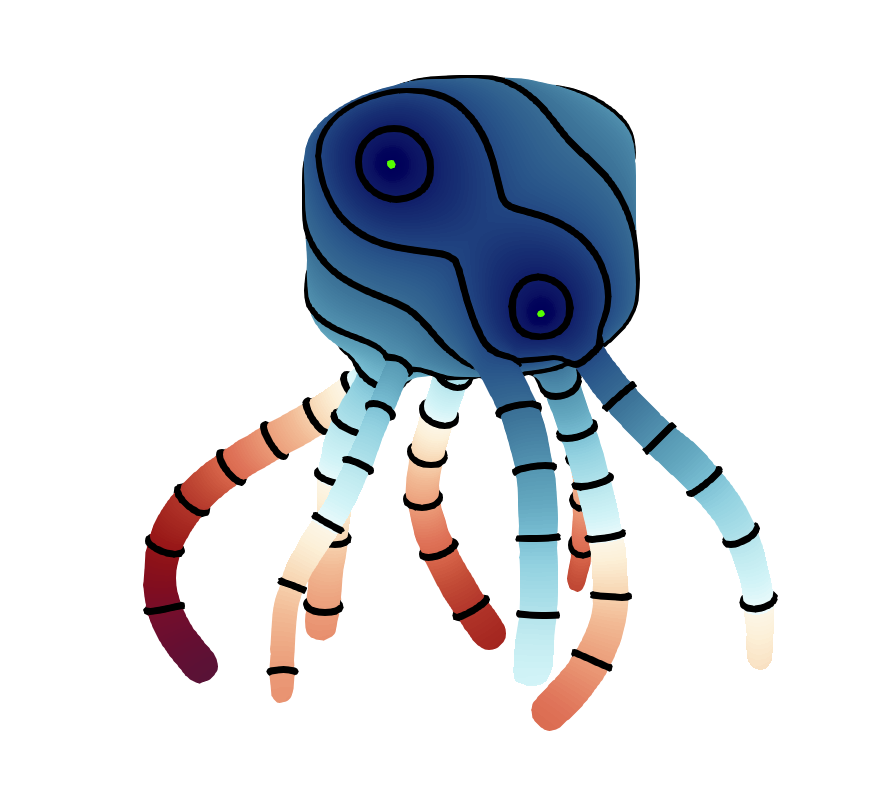}
    \includegraphics[width=0.49\linewidth]{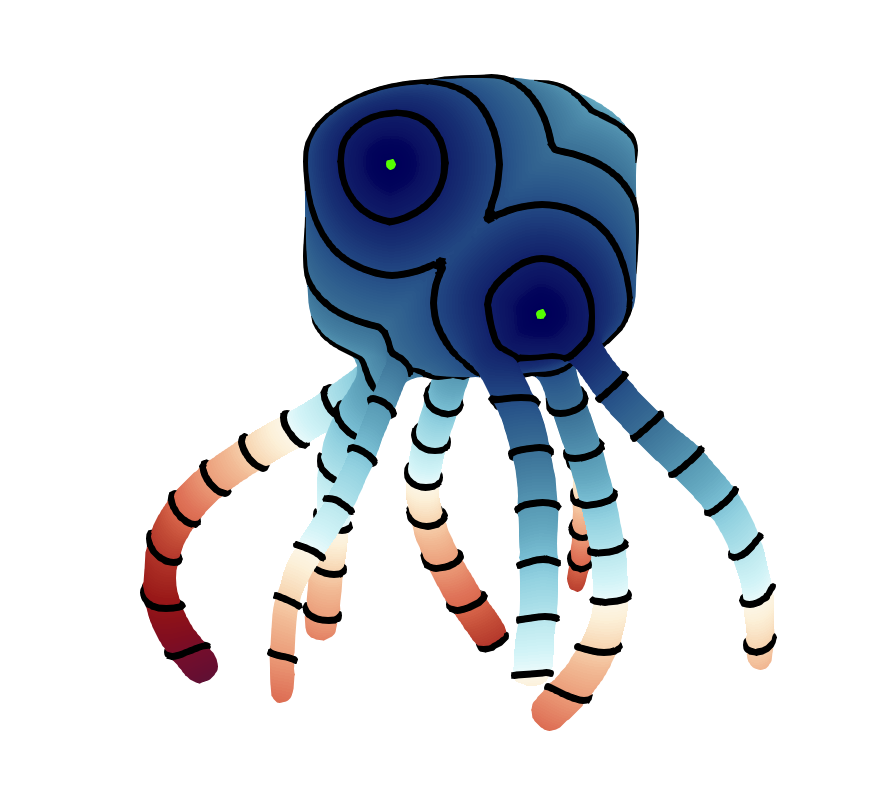}\\

    \includegraphics[width=0.49\linewidth]{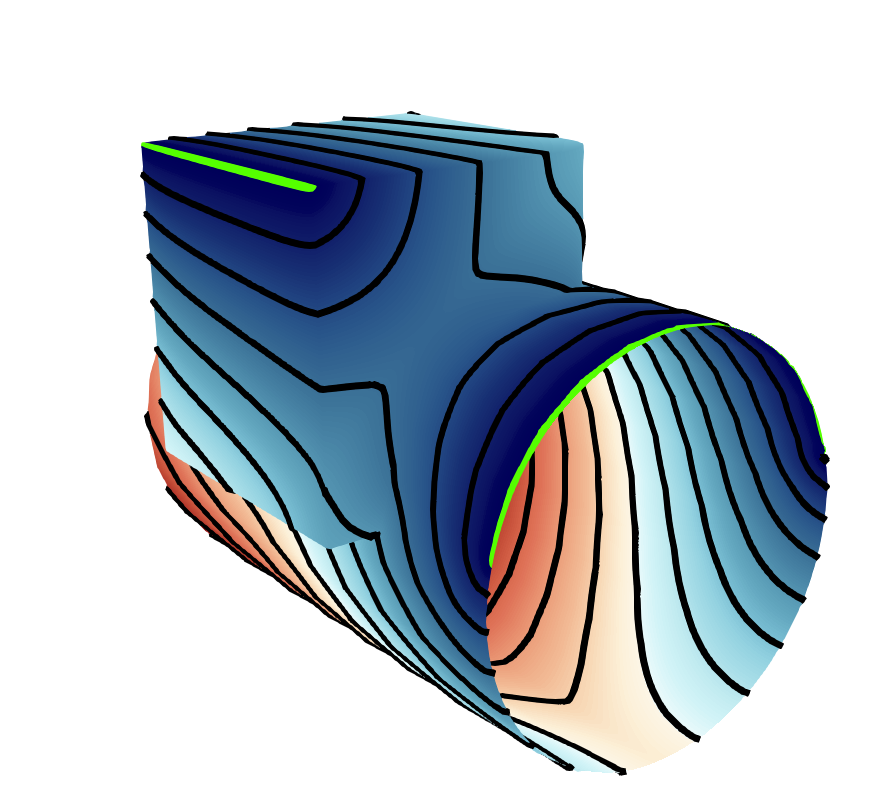}
    \includegraphics[width=0.49\linewidth]{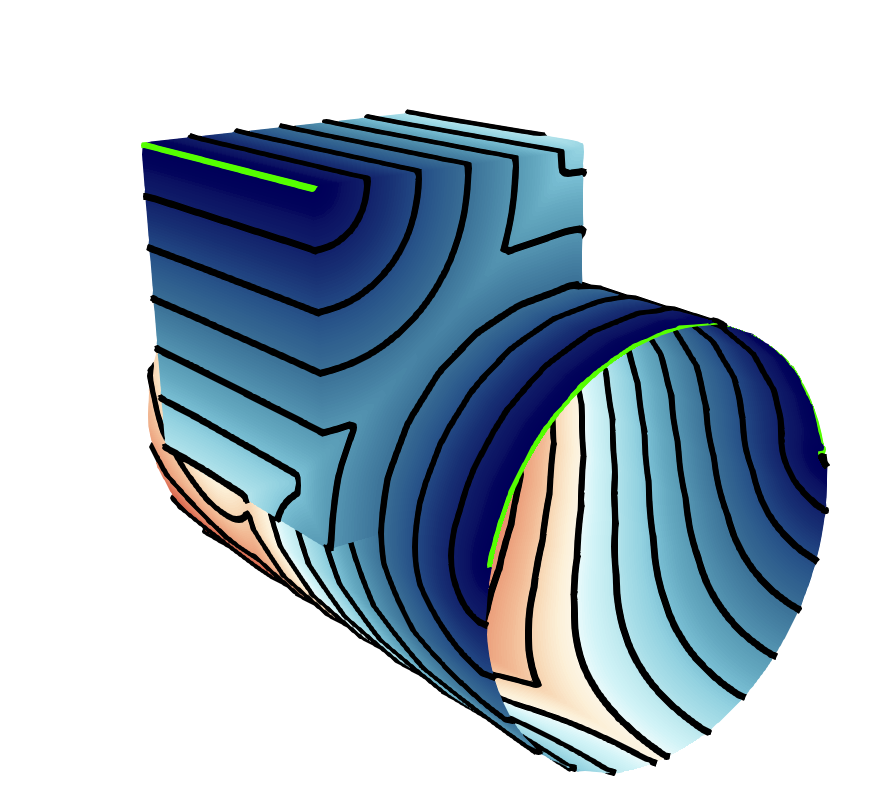}
    \columnbreak \\
    heat distance \cite{Crane} \hspace{2em} polyhedral \cite{Mitchell} \\
    \hrulefill 

    \includegraphics[width=0.45\linewidth]{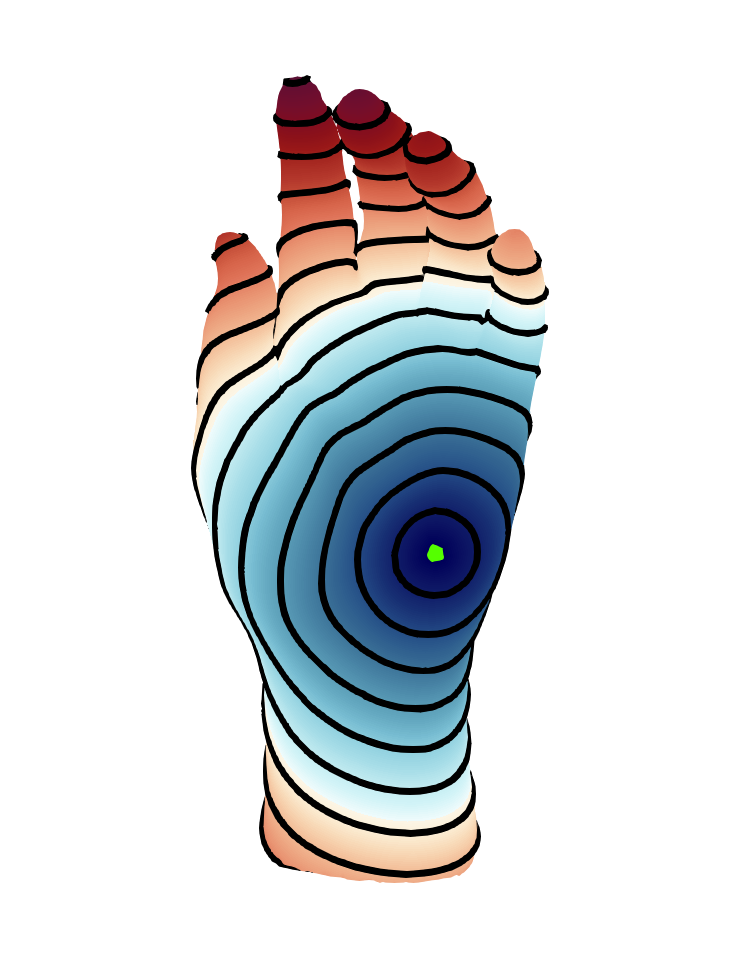}
     \includegraphics[width=0.45\linewidth]{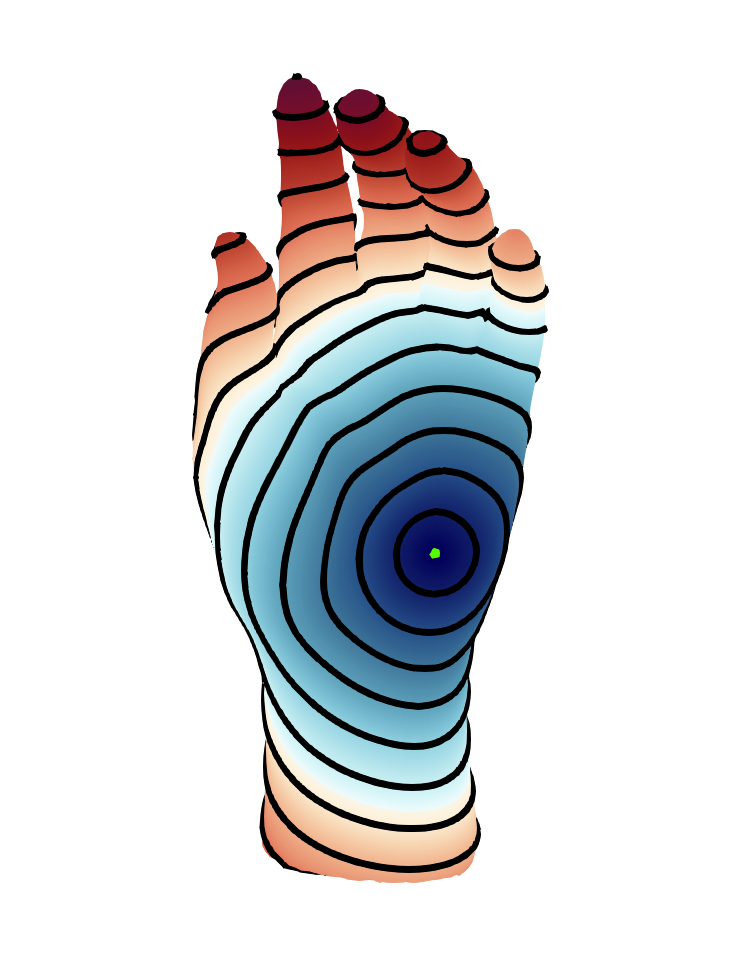}\\ 

    \includegraphics[width=0.49\linewidth]{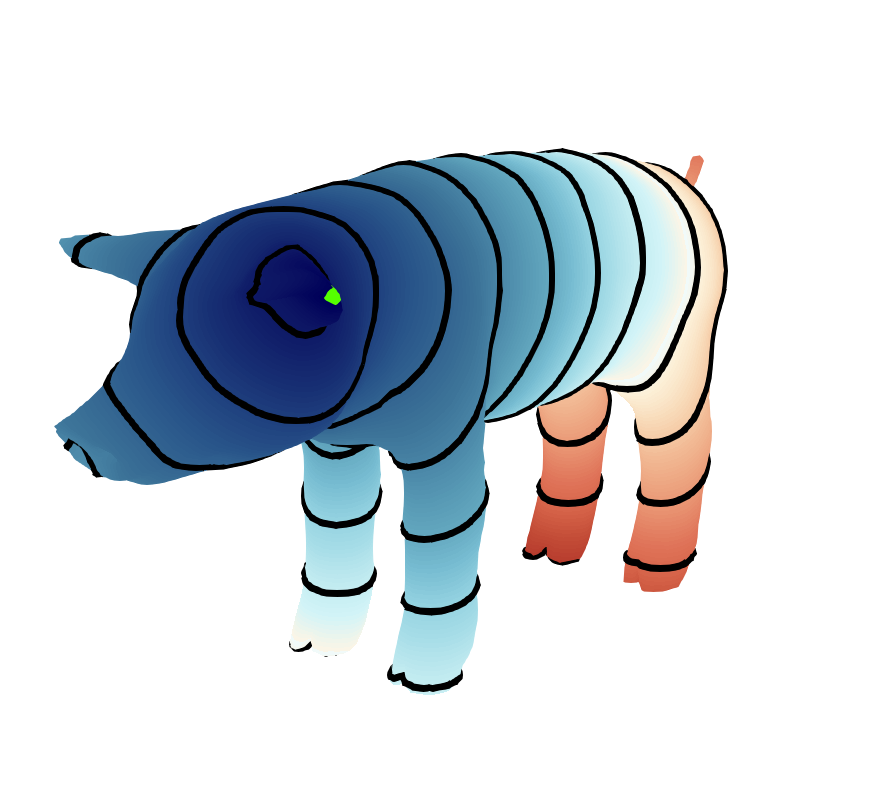}
    \includegraphics[width=0.49\linewidth]{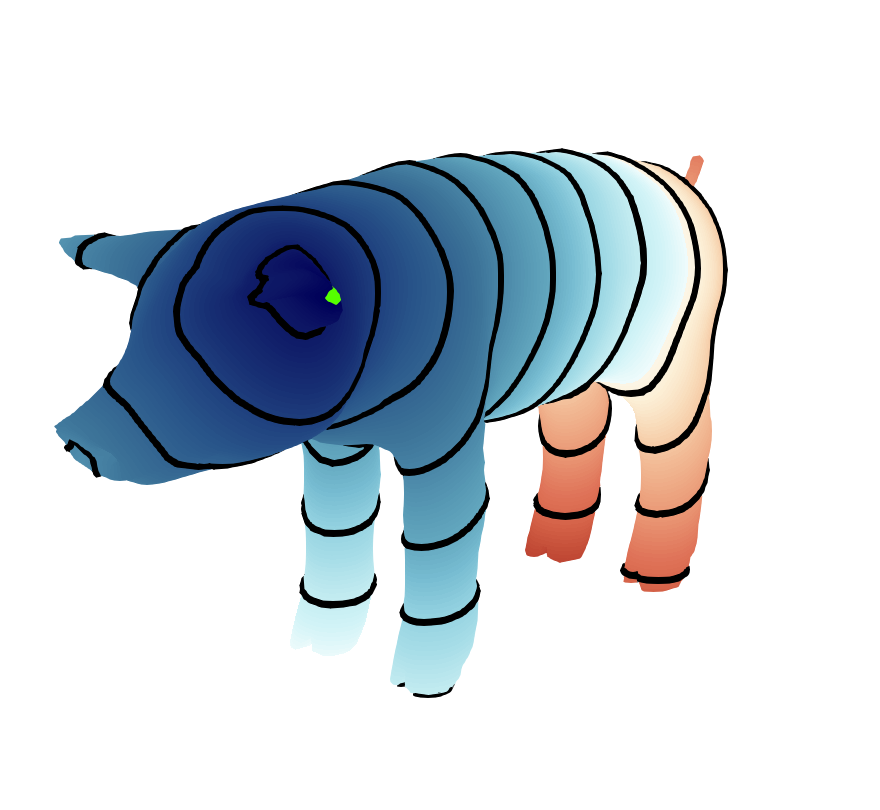}\\

    \includegraphics[width=0.49\linewidth]{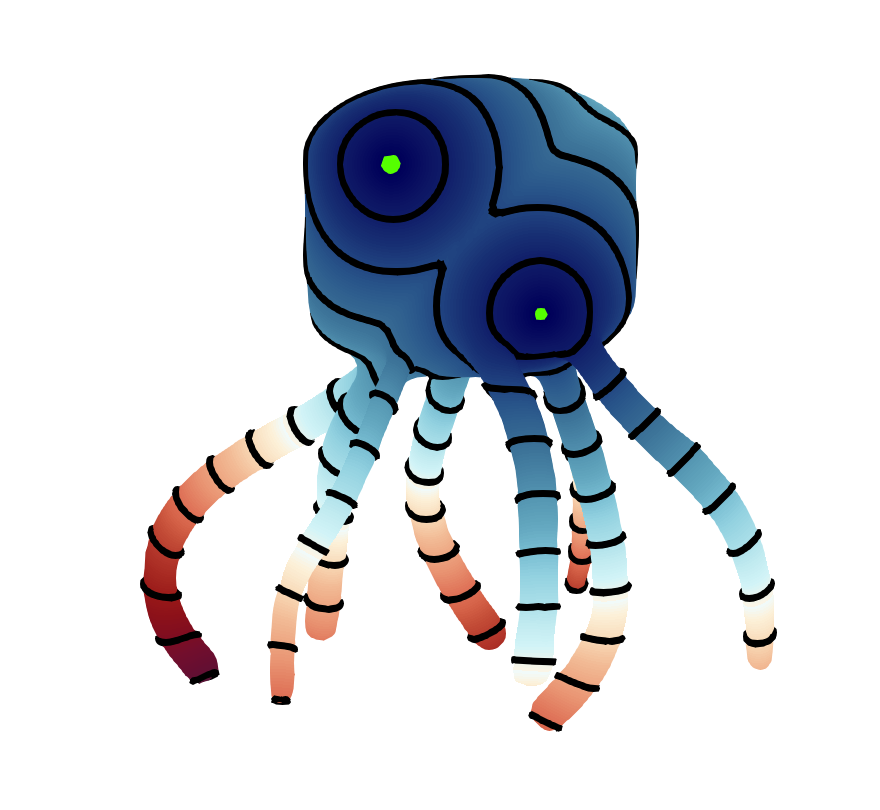}
    \includegraphics[width=0.49\linewidth]{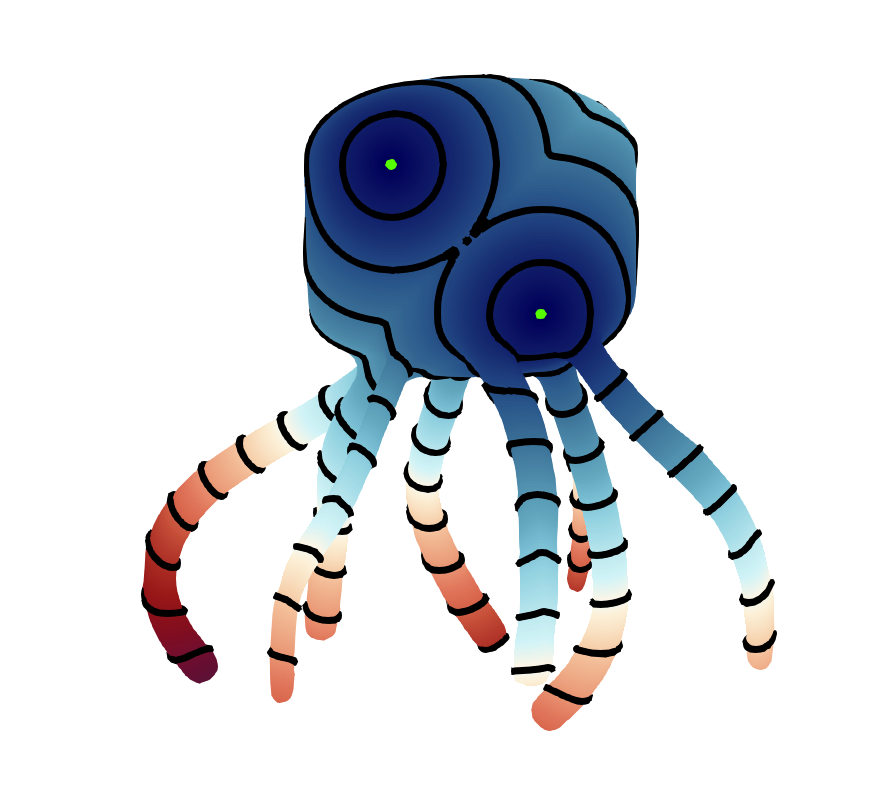}\\

    \includegraphics[width=0.49\linewidth]{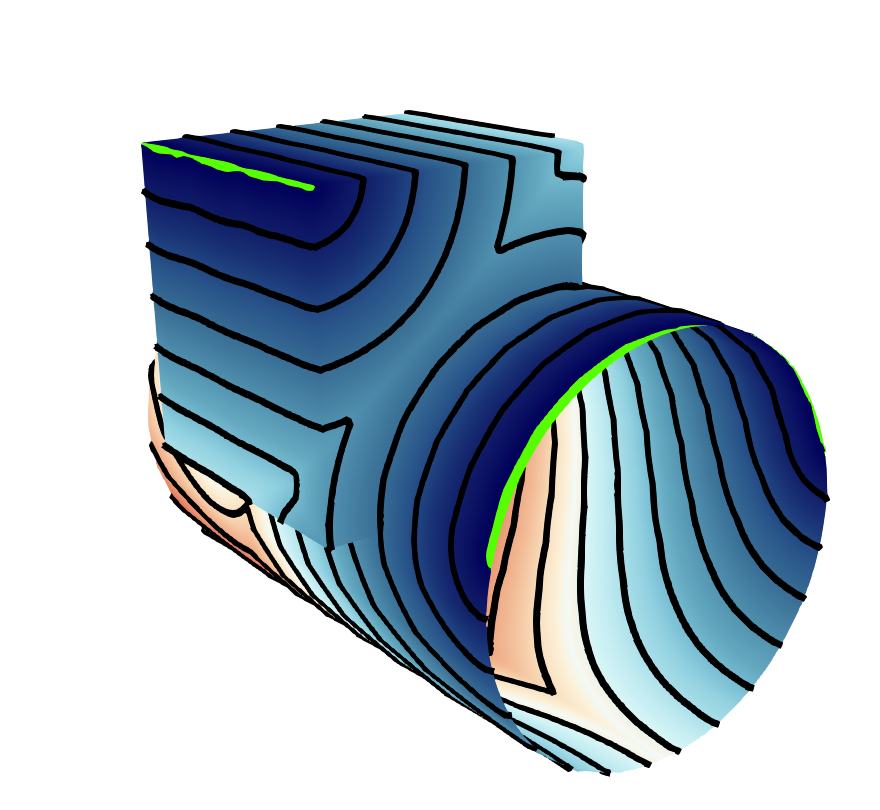}
    \includegraphics[width=0.49\linewidth]{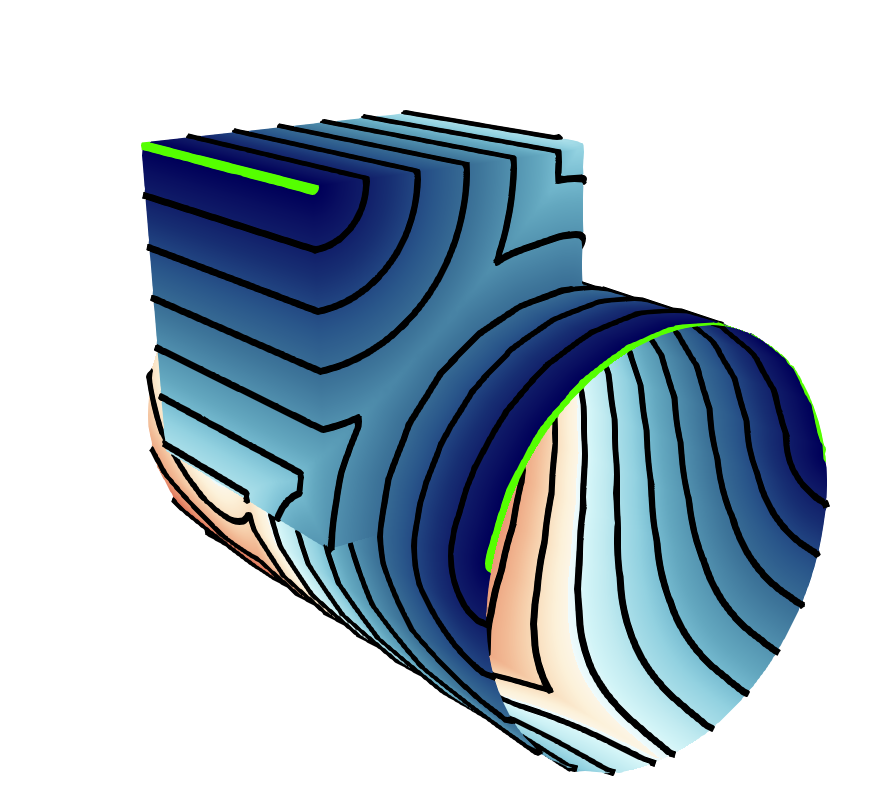}
\end{multicols}
    
    \caption{Computed distance approximations using three methods: the $p$-Poisson method (first and second columns, with $p=5$ and $p=100$, respectively), the \heatmethod (third column), and the polyhedral method (fourth column). For visualization near the feature set $\Gamma_1$, we highlight in green values falling below a chosen tolerance. Top row: distance to a point on the palm of a hand. Second row: distance to the tip of an ear on the pig. Third row: distance to two distinct points on an octopus. Bottom row: distance to two open curves on a surface with boundaries. Each row uses identical surface triangulations. The hand consists of $8647$ vertices and $17290$ faces, the pig consists of $8411$ vertices and $16818$ faces, the octopus consists of $11051$ vertices and $22098$ faces, and the cylinder with box consists of  $81199$ vertices and $161648$ faces. }
    \label{fig:manyEx}
\end{figure}

\Cref{fig:manyEx} shows the computed distance-to-feature $\Gamma_1$ on four different surfaces, using the $p$-Poisson distance approximation with $p=5$ and $p=100$ (first and second columns, respectively), the \heatmethod (third column) and the polyhedral method (fourth column). The top two rows show distances to a point on a closed surface (hand and pig). For both of these examples, the $p=5$ solution overestimates the distance, while the $p=100$ solution shows a more uniform spacing between contours. These differences are most noticeable on the fingers and legs. The \heatmethod and the polyhedral method for these examples give comparable results to our method with $p=100$.

The third row in \Cref{fig:manyEx} shows the distance to a feature set consisting of two points on an octopus. Here, there is a significant increase in sharpness of the contours going from $p = 5$ to $p = 100$. The \heatmethod gives slightly more smoothing, most prominently around the curve which is equidistant to the two points of the feature set -- a similar issue was pointed out for the example in Section \ref{subsect:curve-point} (see the second row of \Cref{fig:ExactExamples}). The polyhedral method gives qualitatively similar contours to the $p$-Poisson method with $p=100$. Lastly, the bottom row shows the distance to two open curves on a surface with boundaries. We have sliced off one side of the cylinder so this example has a nonempty set $\Gamma_2$, as well as sharp edges on the box. Similar to the octopus, we observe a smoother result for $p=5$. The other three methods ($p=100$, heat distance, and polyhedral) give similar contours. 

\subsubsection{Properties of the distance function}
We perform some numerical tests to see how well the $p$-Poisson distance approximations satisfy some fundamental properties of the distance function. Namely, that its gradient has norm $1$ and that distances obey the triangle inequality. 

\begin{figure}
    \centering
    \begin{subfigure}{0.49\textwidth}
    \centering
        \includegraphics[scale = 0.6]{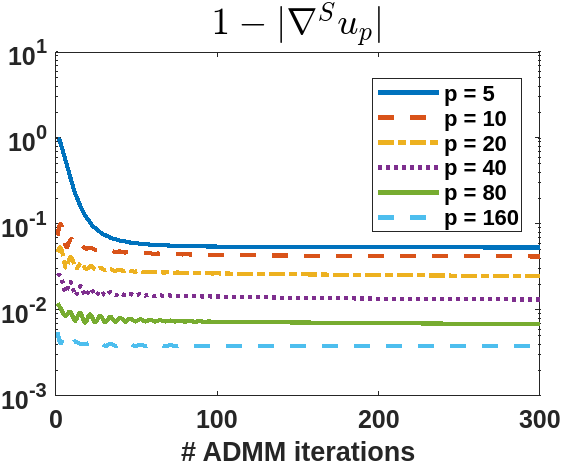}
        \caption{distance to a point on hemisphere} 
    \end{subfigure}
    \begin{subfigure}{0.49\textwidth}        \centering
        \includegraphics[scale = 0.6]{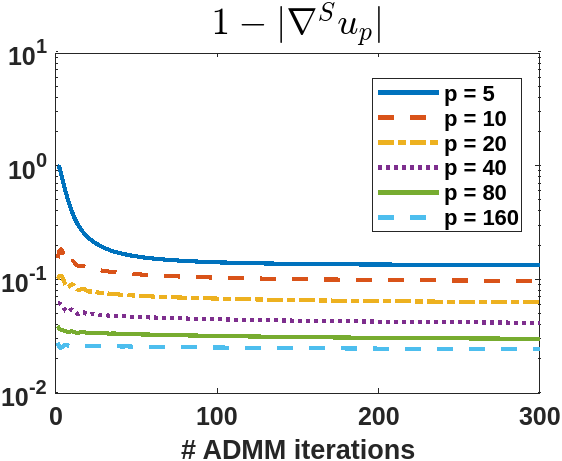}
          \caption{distance to the ear on pig}
    \end{subfigure}
    \caption{Normalized $L^1$ norm of $1-\left| \gradS u_p \right|$ versus ADMM iterations for various $p$-values. Results are shown for the distance to a point on hemisphere in \Cref{fig:ExactExamples} (top row) and the distance to the tip of an ear on the pig in \Cref{fig:manyEx} (second row). Note the convergence to $1$ of $\left| \gradS u_p \right|$ as $p$ increases, indicating that the method converges to the exact geodesic distance. The hemisphere mesh consists of $ 40960$ tri cells with $\ell \approx  \times 10^{-2}, h \approx \times 10^{-2}$ and the pig mesh consists of $16818$ tri cells with $\ell \approx 2.34 \times 10^{-2}, h \approx 4.89 \times 10^{-2}$.
}
    \label{fig:pigGrads}

\end{figure}

\Cref{fig:pigGrads} shows the normalized $L^1$ norm of $1-\left| \gradS u_p \right| $ versus the number of ADMM iterations for various $p$-values. The examples considered in \Cref{fig:pigGrads} are the distance-to-point on the hemisphere, as shown in row 1 of \Cref{fig:ExactExamples}, and the distance-to-ear example on the pig mesh, as shown in row 2 of \Cref{fig:manyEx}. The figure shows that the norm of the computed solution gradient is approaching a value of $1$. The true distance function gradient has norm $1$, so this is an indicator that we are getting closer to the exact distance as $p$ increases. We initialize each of the larger $p$-values ($p \geq 10$) with the solution from the previous $p$-value. Therefore, the most rapid decrease in $1-\left| \gradS u_p \right|$ occurs only during the first few iterations for $p = 5$. This behaviour is also observed for other examples. 

\begin{figure}
    \centering
\begin{subfigure}{0.32\textwidth}
\centering
    \includegraphics[scale = 0.15]{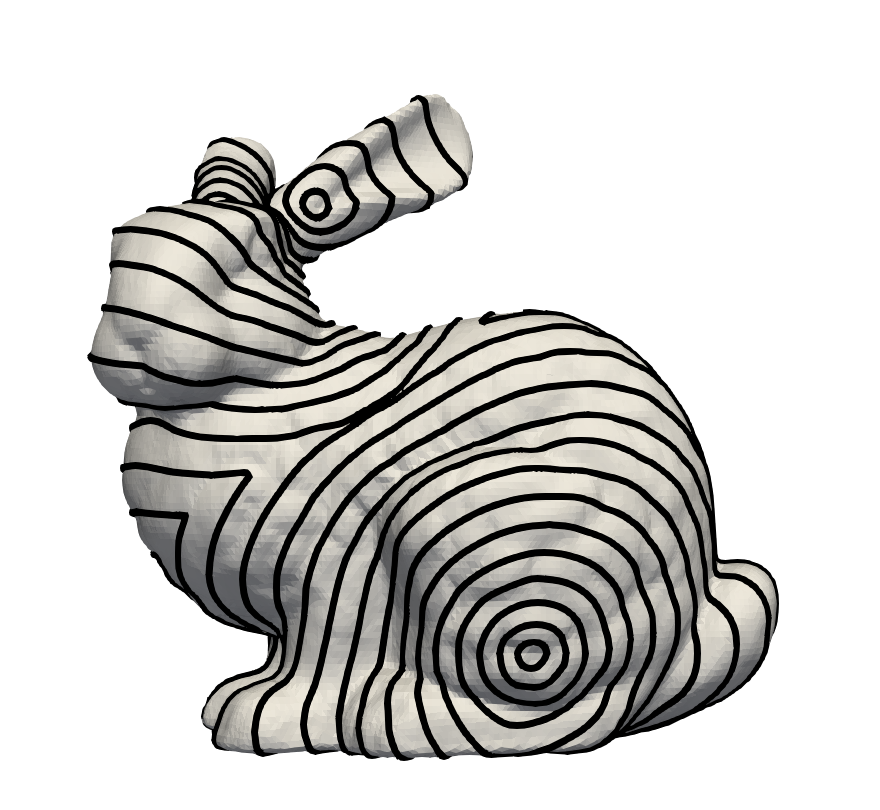}
    \caption{$p$-Poisson with $p = 5$}
\end{subfigure}
\begin{subfigure}{0.32\textwidth}
\centering
    \includegraphics[scale = 0.15]{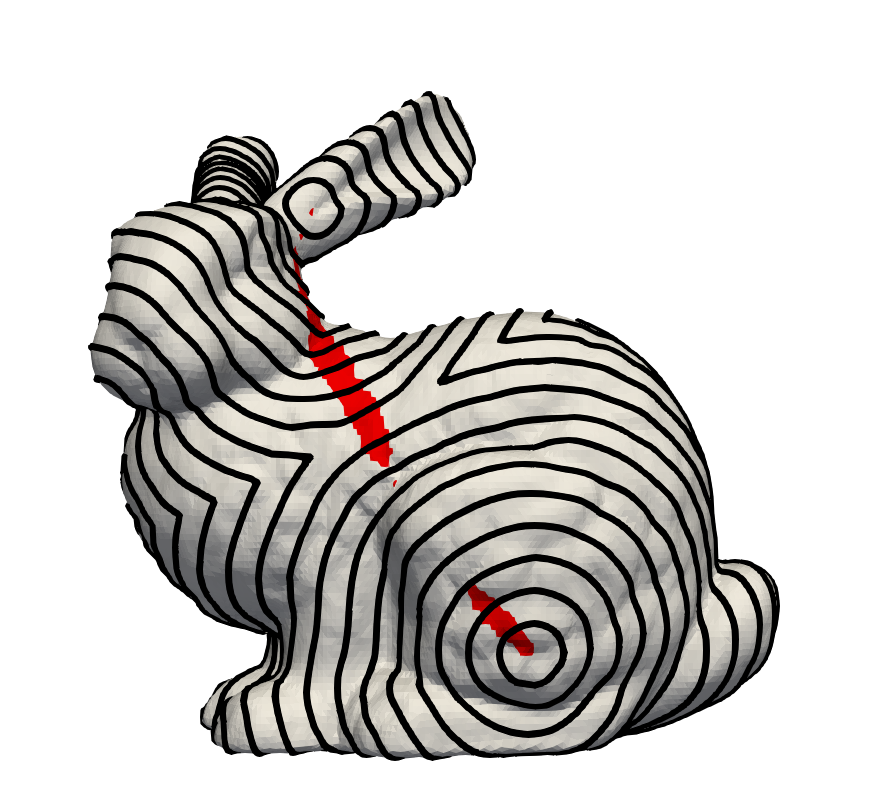}
    \caption{heat distance \cite{Crane} }
\end{subfigure}
\begin{subfigure}{0.32\textwidth}
\centering
    \includegraphics[scale = 0.15]{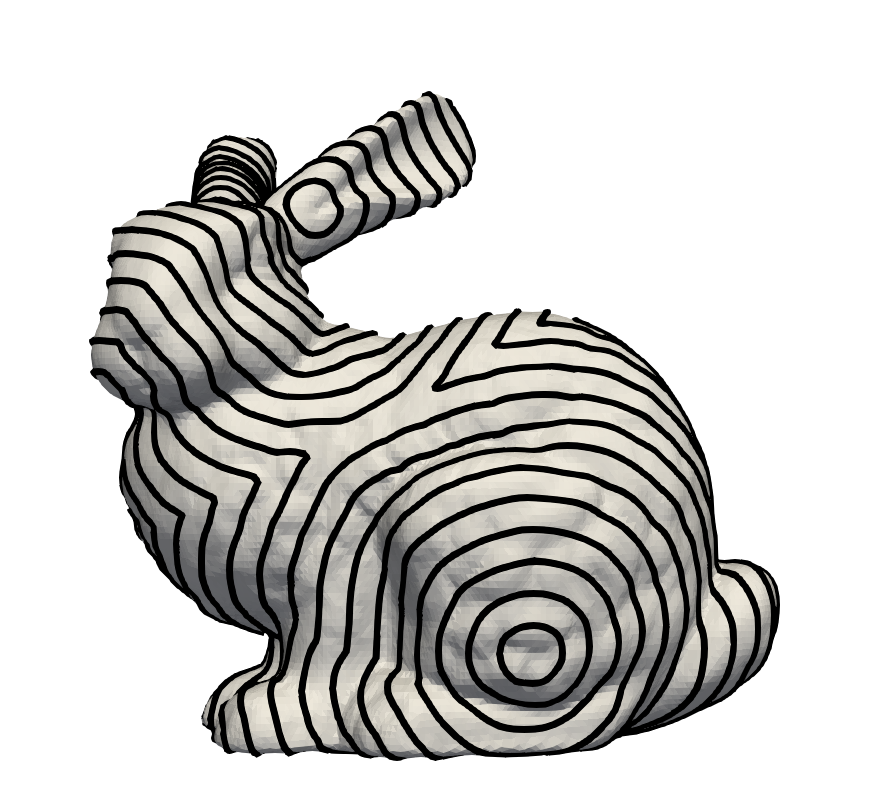}
    \caption{polyhedral \cite{Mitchell} }
\end{subfigure}

    \caption{Triangle inequality check. The feature set consists of a point $q_1$ on the ear on the bunny, and a point $q_2$ on its lower body. The computations were done with three methods: our $p$-Poisson distance approximation (left), the \heatmethod (center), and the polyhedral method (right). Regions which satisfy the triangle inequality are shown in white, along with the level sets of $\min\{d(q_1, x), d(q_2, x)\}$, where $d(\cdot, x)$ denotes the distance approximation obtained by the method of consideration. Regions which violate the triangle inequality are shown in red. Note that the entire bunny is white for the $p$-Poisson distance approximation and the polyhedral method, while the \heatmethod fails to satisfy the triangle inequality on fairly large regions. The bunny triangulation consists of $3900$ vertices and $7800$ faces.
}
    \label{fig:triangle-inequality}
\end{figure}

\Cref{fig:triangle-inequality} illustrates how well various numerical methods for computing geodesic distances satisfy the triangle inequality. The figure shows computations of the distance on a bunny, where the feature set consists of a point $q_1$ on the ear and a point $q_2$ on the lower body. The left plot was obtained with our method (using $p = 5$), the middle plot used the heat method, and the one on the right corresponds to the polyhedral method. We note that there is a small difference between our computations of $d_p(q_1, q_2)$ and $d_p(q_2, q_1)$, and for this reason we test the triangle inequality using the maximum of the two distances, i.e., we check
\[
d_p(q_1, x) + d_p(q_2, x) \geq \max\{ d_p(q_1, q_2), d_p(q_2, q_1)\}, \quad \text{ for } x \in S.
\]
In \Cref{fig:triangle-inequality}, the regions that satisfy the triangle inequality above are shown in white, along with the level sets of $\min\{d(q_1, x), d(q_2, x)\}$. Here, $d(\cdot, x)$ denotes the approximate distance to $x$ computed using the method under consideration. The figure shows that throughout the entire mesh, both our method and the polyhedral method satisfy the inequality. We also point out that our method has achieved this result with a low value of $p$, and the inequality remains satisfied for all higher values of $p$ that we tested. In contrast, distances computed using the \heatmethod fail to satisfy the triangle inequality over significant regions on the bunny, shown in red. This shortcoming of the \heatmethod was already noted in \cite{Solomon} (see Figure~9 in their paper for another visualization of these regions).

\subsubsection{Robustness against mesh quality}
\label{sec:robustness}

It is desirable for a method to withstand some topological perturbations such as vertex noise, artificial holes, and nearly isometric deformation.

\begin{figure}
    \centering

    \begin{subfigure}{0.49\textwidth}
    \centering
    \includegraphics[scale = 0.12]{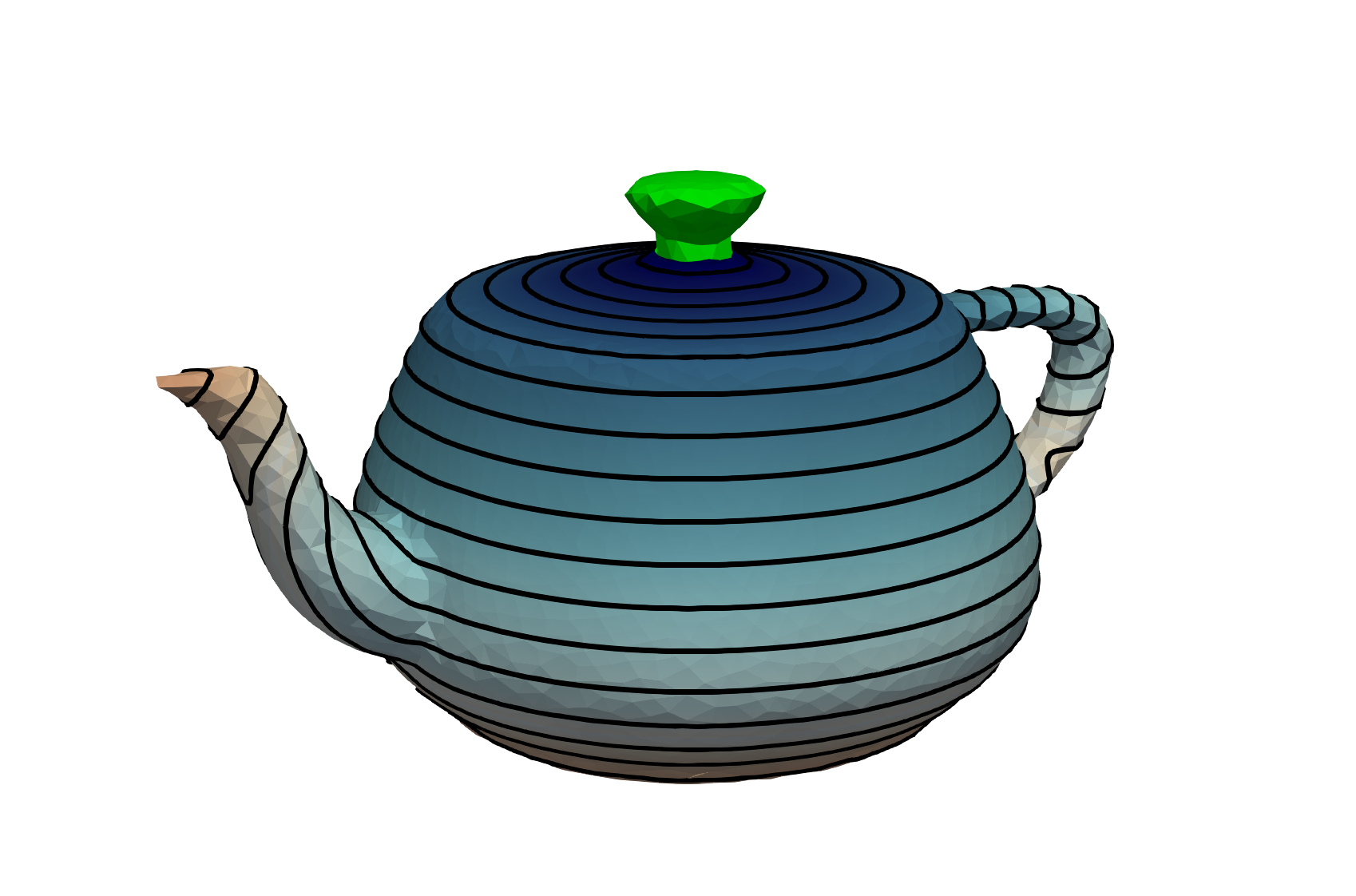} 
     \caption{$\sigma = 0$}
    \end{subfigure}
    \begin{subfigure}{0.49\textwidth}
    \centering
    \includegraphics[scale = 0.12]{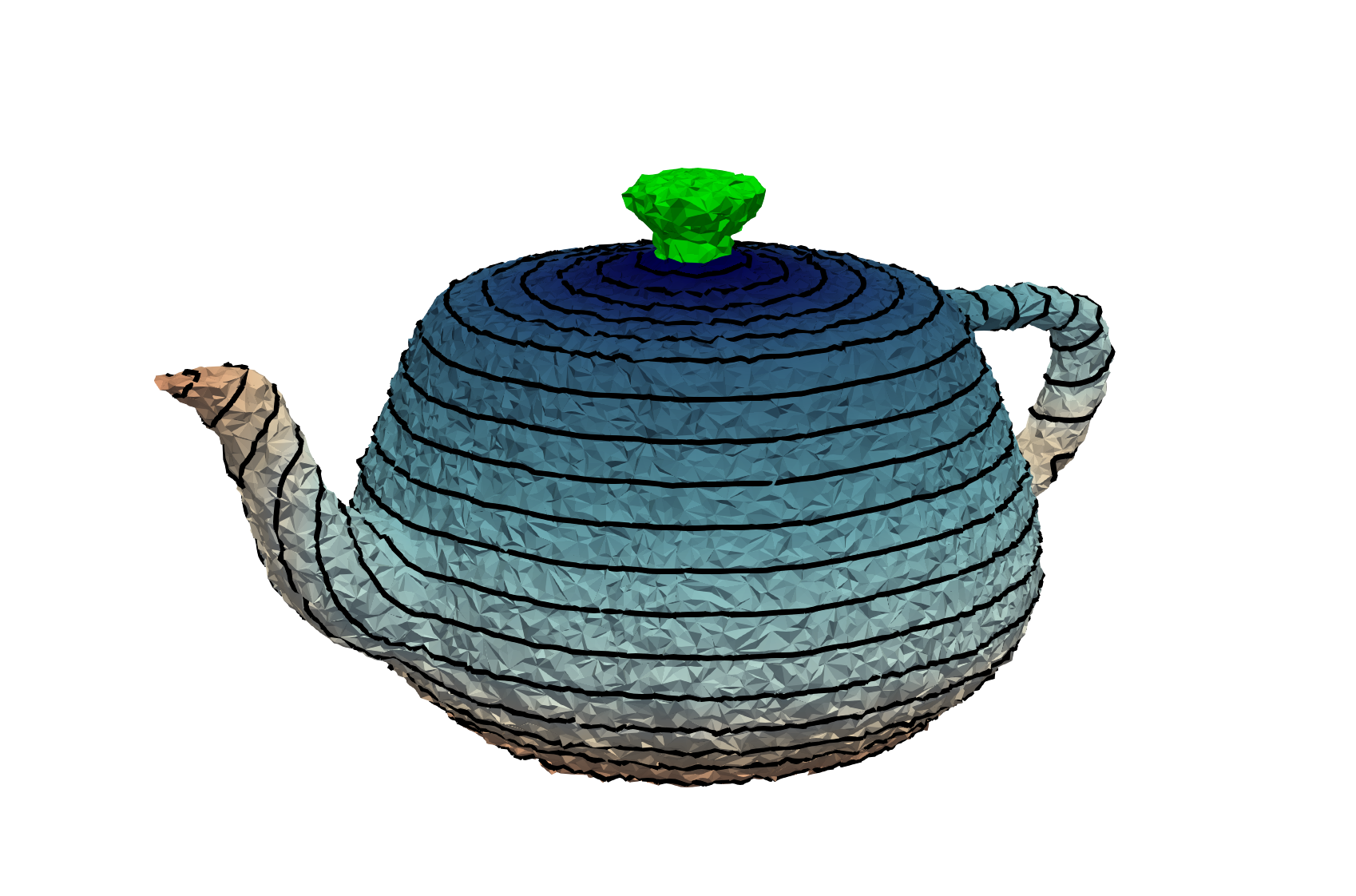} 
     \caption{$\sigma = \frac{\ell}{8}$}
    \end{subfigure}
    \begin{subfigure}{0.49\textwidth}
    \centering
    \includegraphics[scale = 0.12]{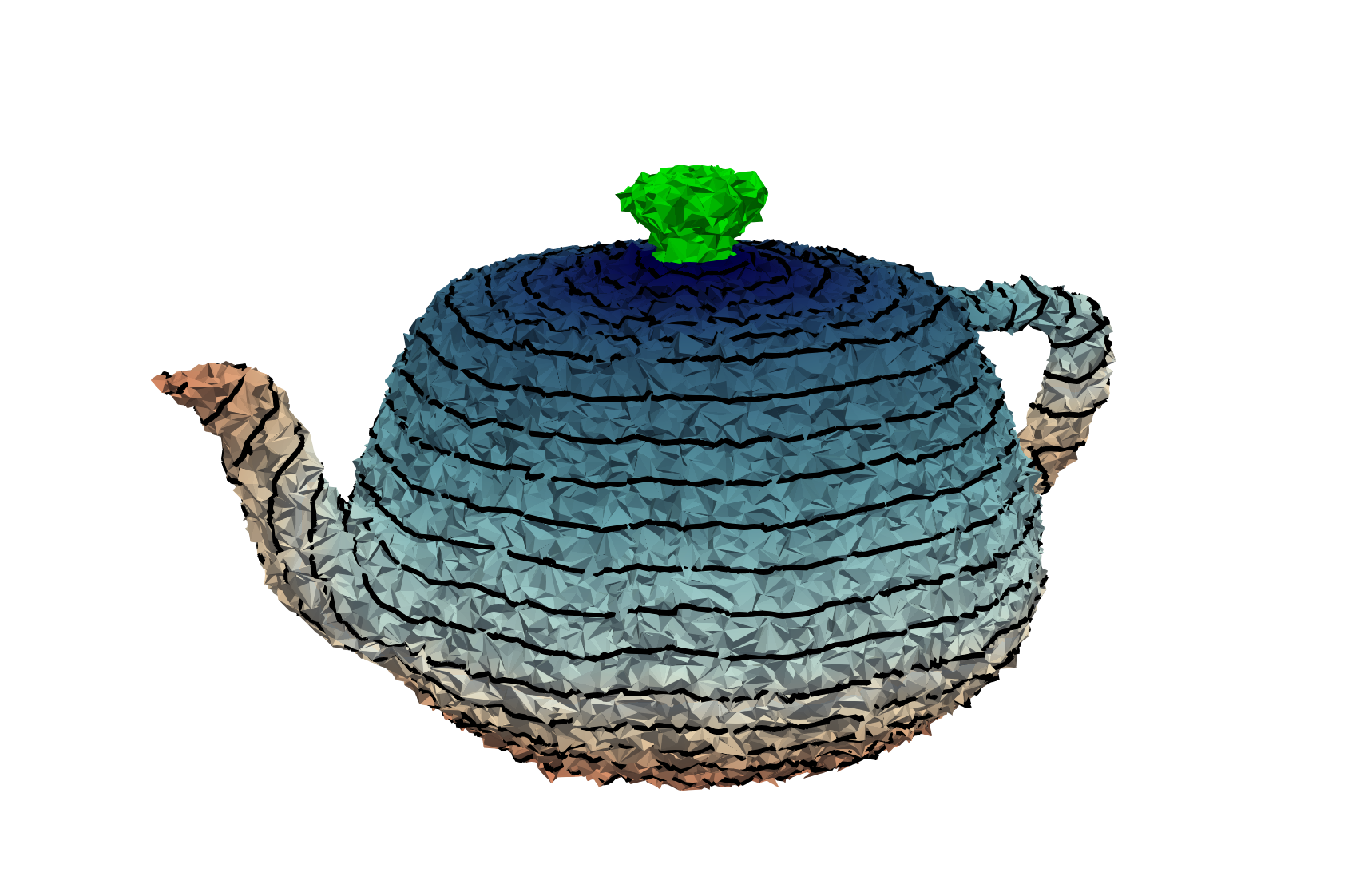} 
     \caption{$\sigma = \frac{\ell}{4}$}
    \end{subfigure}
    \begin{subfigure}{0.49\textwidth}
    \centering
    \includegraphics[scale = 0.12]{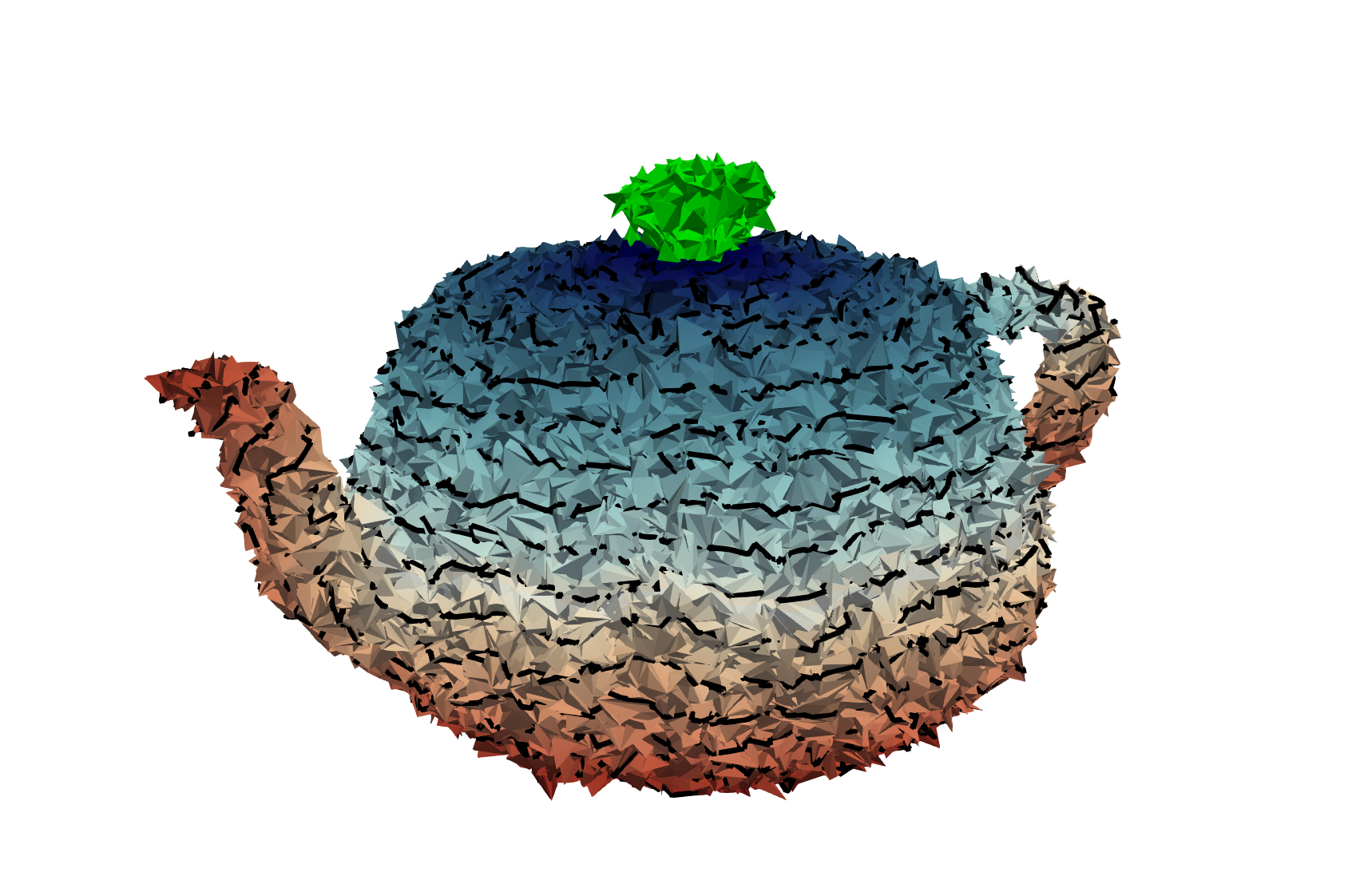} 
    \caption{$\sigma = \frac{\ell}{2}$}
    \end{subfigure}

    \caption{Computed $p$-Poisson distances to the lid of a teapot with Gaussian noise added to the mesh vertices. The standard deviation is in terms of the average mesh cell edge length $\ell \approx  1.60 \times 10^{-2}$. Top left: no noise. Top right: minimal amount of noise ($\sigma = \frac{\ell}{8}$). Bottom left: medium amount of noise ($\sigma = \frac{\ell}{4}$) . Bottom right: large amount of noise ($\sigma = \frac{\ell}{2}$). The teapot quad mesh consists of $3900$ vertices and $7800$ faces. }
    \label{fig:noise}
\end{figure}

Crane et al. \cite{Crane} and Solomon et al. 
\cite{Solomon} both demonstrate that their methods for distance approximation can withstand some per-vertex noise.  \Cref{fig:noise} shows computed $p$-Poisson distances to the lid of a teapot with varying degrees of Gaussian noise added to the mesh. More precisely, we perturb the vertices by adding Gaussian noise with standard deviations $\sigma = \frac{\ell}{8}, \frac{\ell}{4}, \frac{\ell}{2}$ where $\ell \approx  1.60 \times 10^{-2}$ is the average cell edge length on the mesh.  We observe that the contours undergo minimal change with the addition of noise. Visualizations are shown only for $p = 100$ although other $p$-values have similar results. 

\section{Conclusions}
\label{sec:conclusions}

We have extended the $p$-Poisson distance problem to handle the intrinsic distance-to-feature problem. Our approach has the advantage that errors are controlled near the surface boundary without the need to average multiple computations, as is required with Crane et al.'s \cite{Crane} method. The ADMM scheme for solving the boundary value problem demonstrates numerical convergence to the exact geodesic distance as $p \to \infty$. Additionally, ADMM decouples the ill-conditioning effects of large $p$-values, which means the starting $p$-value may be large. Our computed $p$-Poisson distances exhibit many desirable properties, such as robustness against geometric noise and satisfaction of the triangle inequality. Our workflows are centered on the deal.II implementation for triangular and quadrilateral meshes, leaving room for future work exploring $p$-Poisson distances on various discrete surface representations, as well as numerical approximation methods. The numerical methods presented are also adaptable to different problems involving $\Delta ^S_p$, so this work provides a starting point for the study of other applications. 

\section*{Acknowledgements}  
\label{sec:acknowledgements} 
The authors gratefully acknowledge the financial support of NSERC Canada (RGPIN 341834 for RF and RGPIN 03302 for SR).
 
The bunny mesh is provided courtesy of the Stanford Computer Graphics Laboratory. The remaining meshes are obtained from the Chen et al. \cite{offMESHES} benchmark set for segmentation.




 \bibliographystyle{elsarticle-num} 
 \bibliography{main}

\begin{thebibliography}{10}
\expandafter\ifx\csname url\endcsname\relax
  \def\url#1{\texttt{#1}}\fi
\expandafter\ifx\csname urlprefix\endcsname\relax\def\urlprefix{URL }\fi
\expandafter\ifx\csname href\endcsname\relax
  \def\href#1#2{#2} \def\path#1{#1}\fi

\bibitem{FAYOLLE20181}
P.-A. Fayolle, A.~G. Belyaev, p-{L}aplace diffusion for distance function estimation, optimal transport approximation, and image enhancement, Computer Aided Geometric Design 67 (2018) 1--20.

\bibitem{Crane}
K.~Crane, C.~Weischedel, M.~Wardetzky, The heat method for distance computation, Commun. Association for Computing Machinery 60~(11) (2017) 90--99.

\bibitem{Mitchell}
J.~S.~B. Mitchell, D.~M. Mount, C.~H. Papadimitriou, The discrete geodesic problem, Society for Industrial and Applied Mathematics J. Comput. 16 (1987) 647--668.

\bibitem{BuhlerHein}
T.~B\"{u}hler, M.~Hein, Spectral clustering based on the graph p-{L}aplacian, in: Proceedings of the 26th Annual International Conference on Machine Learning, Association for Computing Machinery, 2009, p. 81–88.

\bibitem{Blomgren}
P.~Blomgren, T.~Chan, P.~Mulet, C.~Wong, Total variation image restoration: Numerical methods and extensions, in: Institute of Electrical and Electronics Engineers International Conference on Image Processing, Vol.~3, 1997, pp. 384 -- 387.

\bibitem{Chen2006}
Y.~Chen, S.~Levine, M.~Rao, Variable exponent, linear growth functionals in image restoration, Society for Industrial and Applied Mathematics Journal on Applied Mathematics 66~(4) (2006) 1383--1406.

\bibitem{Caselles}
V.~Caselles, L.~Igual, O.~Sander, An axiomatic approach to scalar data interpolation on surfaces, Numerische Mathematik 102 (2006) 383--411.

\bibitem{Ruzicka}
M.~Ruzicka, Electrorheological fluids: Modeling and mathematical theory, Lecture Notes in. Mathematics, Springer Berlin Heidelberg 1748 (2000).

\bibitem{shapeMet}
G.~Cong, M.~Esser, B.~Parvin, G.~Bebis, Shape metamorphism using p-{L}aplacian equation, in: Proceedings of the 17th International Conference on Pattern Recognition, 2004. ICPR 2004., Vol.~4, 2004, pp. 15--18.

\bibitem{bhattacharya1989limits}
T.~Bhattacharya, E.~DiBenedetto, J.~Manfredi, Limits as $p \to \infty$ of {$\Delta_p u = f$} and related extremal problems, Rend. Sem. Mat. Univ. Politec. Torino 47 (1989) 15--68.

\bibitem{Lipman}
Y.~Lipman, R.~M. Rustamov, T.~A. Funkhouser, Biharmonic distance, Association for Computing Machinery Trans. Graph. 29~(3) (2010).

\bibitem{Solomon}
J.~Solomon, R.~Rustamov, L.~Guibas, A.~Butscher, Earth mover's distances on discrete surfaces, Association for Computing Machinery Trans. Graph. 33~(4) (2014).

\bibitem{NaberMedImage}
A.~Naber, D.~Berwanger, W.~Nahm, Geodesic length measurement in medical images: Effect of the discretization by the camera chip and quantitative assessment of error reduction methods, Photonics 7~(3) (2020).

\bibitem{Belyaev2015}
A.~G. Belyaev, P.-A. Fayolle, On variational and {PDE}-based distance function approximations, Computer Graphics Forum 34~(8) (2015) 104--118.

\bibitem{WANG2017262}
X.~Wang, Z.~Fang, J.~Wu, S.-Q. Xin, Y.~He, Discrete geodesic graph ({DGG}) for computing geodesic distances on polyhedral surfaces, Computer Aided Geometric Design 52-53 (2017) 262--284.

\bibitem{KimmelSethian1998}
R.~Kimmel, J.~A. Sethian, Computing geodesic paths on manifolds, Proceedings of the National Academy of Sciences 95~(15) (1998) 8431--8435.

\bibitem{Varadhan}
S.~R.~S. Varadhan, Diffusion processes in a small time interval, Communications on Pure and Applied Mathematics 20~(4) (1967) 659--685.

\bibitem{FengCrane}
N.~Feng, K.~Crane, A heat method for generalized signed distance, ACM Trans. Graph. 43~(4) (jul 2024).

\bibitem{Dijkstra1959}
E.~W. Dijkstra, A note on two problems in connexion with graphs, Numerische Mathematik 1~(1) (1959) 269--271.

\bibitem{crane2020survey}
K.~Crane, M.~Livesu, E.~Puppo, Y.~Qin, A survey of algorithms for geodesic paths and distances, arXiv preprint arXiv:2007.10430 (2020).

\bibitem{Sethian1996}
J.~A. Sethian, A fast marching level set method for monotonically advancing fronts., Proceedings of the National Academy of Sciences 93~(4) (1996) 1591--1595.

\bibitem{ConvexNew}
M.~Edelstein, N.~Guillen, J.~Solomon, M.~Ben-Chen, A convex optimization framework for regularized geodesic distances, in: Association for Computing Machinery SIGGRAPH 2023 Conference Proceedings, SIGGRAPH '23, Association for Computing Machinery, 2023.

\bibitem{Adikusuma}
Y.~Y. Adikusuma, Z.~Fang, Y.~He, Fast construction of discrete geodesic graphs, ACM Trans. Graph. 39~(2) (Mar. 2020).

\bibitem{dealII94}
D.~Arndt, W.~Bangerth, M.~Feder, M.~Fehling, R.~Gassm{\"o}ller, T.~Heister, L.~Heltai, M.~Kronbichler, M.~Maier, P.~Munch, J.-P. Pelteret, S.~Sticko, B.~Turcksin, D.~Wells, The \texttt{deal.II} library, version 9.4, Journal of Numerical Mathematics 30~(3) (2022) 231--246.

\bibitem{Stein}
O.~Stein, E.~Grinspun, M.~Wardetzky, A.~Jacobson, Natural boundary conditions for smoothing in geometry processing, ACM Transactions on Graphics 37~(2), article 23 (2018).

\bibitem{Azorero_etal2009-mlimits}
J.~Garcia-Azorero, J.~J. Manfredi, I.~Peral, J.~D. Rossi, The limit as $p \to \infty$ for the $p$-{L}aplacian with mixed boundary conditions and the mass transport problem through a given window, Rend. Lincei Mat. Appl. 20 (2009) 111--126.

\bibitem{Huang2007PreconditionedDA}
Y.~Q. Huang, R.~Li, W.~Liu, Preconditioned descent algorithms for p-{L}aplacian, Journal of Scientific Computing 32 (2007) 343--371.

\bibitem{Aubin-book}
T.~Aubin, Nonlinear Analysis on Manifolds. {M}onge-{A}mp\`ere Equations, 1st Edition, Vol. 252 of Grundlehren der mathematischen Wissenschaften, Springer New York, NY, 1982.

\bibitem{DziukElliott}
G.~Dziuk, C.~M. Elliott, Finite element methods for surface {PDE}s, Acta Numerica 22 (2013) 289–396.

\bibitem{CPM}
S.~J. Ruuth, B.~Merriman, A simple embedding method for solving partial differential equations on surfaces, Journal of Computational Physics 227~(3) (2008) 1943--1961.

\bibitem{ADMM}
S.~Boyd, N.~Parikh, E.~Chu, B.~Peleato, J.~Eckstein, Distributed optimization and statistical learning via the alternating direction method of multipliers, Foundations and Trends in Machine Learning 3 (2011) 1--122.

\bibitem{PDEADMM}
D.~Gabay, B.~Mercier, A dual algorithm for the solution of nonlinear variational problems via finite element approximation, Computers \& Mathematics with Applications 2~(1) (1976) 17--40.

\bibitem{Glowinski1974OnTS}
R.~Glowinski, A.~Marrocco, On the Solution of a Class of Non-Linear Dirichlet Problems by a Penalty-Duality Method and Finite Elements of Order One, Springer Berlin Heidelberg, 1975, pp. 327--333.

\bibitem{paraview}
U.~Ayachit, The ParaView Guide: A Parallel Visualization Application, Kitware, Inc., 2015.

\bibitem{geometrycentral}
N.~Sharp, K.~Crane, et~al., Geometrycentral: A modern c++ library of data structures and algorithms for geometry processing (2019).

\bibitem{Nguyen2019}
T.~Nguyen, T.~Nguyen, B.~M. Nguyen, G.~Nguyen, Efficient time-series forecasting using neural network and opposition-based coral reefs optimization, International Journal of Computational Intelligence Systems 12 (2019) 1144--1161.

\bibitem{offMESHES}
X.~Chen, A.~Golovinskiy, T.~Funkhouser, A benchmark for {3D} mesh segmentation, Association for Computing Machinery Transactions on Graphics (Proc. SIGGRAPH) 28~(3) (2009).

\end{thebibliography}





\end{document}